\documentclass[preprint,showpacs,preprintnumbers,amsmath,amssymb,showkeys]{revtex4}
\usepackage[utf8]{inputenc}
\usepackage[T2A]{fontenc}
\usepackage[ukrainian,english]{babel}
\usepackage{amsthm}
\usepackage{siunitx}  % Для одиниць вимірювання
\usepackage{amssymb}
\usepackage{amsmath}  % для математичних символів
\usepackage{amsfonts}  % для шрифтів з математичними символами

\usepackage{amsmath}
\usepackage{amssymb}
\usepackage{amsmath,amssymb,amsthm,physics}

\usepackage[a4paper,margin=2cm]{geometry}
\usepackage{setspace}
\usepackage{amsmath,amssymb,mathtools,bm,physics,mathrsfs,dsfont}

\usepackage{booktabs,tabularx,makecell,enumitem}

\usepackage{graphicx,wrapfig,subcaption,float,xcolor}
\usepackage{pgfplots,tikz}
\usetikzlibrary{arrows.meta,angles,quotes,calc,decorations.pathmorphing}
\pgfplotsset{compat=1.18}
\usepackage{caption}

\usepackage[colorlinks=true,linkcolor=blue,urlcolor=teal,citecolor=magenta]{hyperref}

\usepackage[most]{tcolorbox}
\allowdisplaybreaks

\tcbset{
  colback=white,
  colframe=black,
  boxrule=0.8pt,
  arc=4pt,
  left=3mm,right=3mm,top=2mm,bottom=2mm,
  before skip=6pt, after skip=8pt,
  fonttitle=\bfseries,
  title style={fill=green!60!black, font=\bfseries\color{white}},
  enhanced,
}

\newcounter{task}
\renewcommand{\thetask}{\arabic{task}}
\newtcolorbox{taskbox}[2][]{title={Задача \thetask. #2}, #1}

\newcommand{\bel}[1]{\begin{equation}\label{#1}}

\def\be{\begin{equation}}
\def\ee{\end{equation}}
\def\bea{\begin{eqnarray}}
\def\eea{\end{eqnarray}}

\def\ms{\medskip}
\def\siml{\;\hbox{\kern.1em \lower.7ex \hbox{$\sim$} \kern-1.12em
\raise.5ex \hbox{$<$} \kern.1em}}
\def\simg{\;\hbox{\kern.1em \lower.7ex \hbox{$\sim$} \kern-1.12em
\raise.5ex \hbox{$>$} \kern.1em}}

\def\siml{\hbox{\kern.1em \lower.6ex \hbox{$\sim$} \kern-1.12em
          \raise.6ex \hbox{$<$} \kern.1em }}
\def\simg{\hbox{\kern.1em \lower.6ex \hbox{$\sim$} \kern-1.12em
          \raise.6ex \hbox{$>$} \kern.1em }}

\ms

\newcommand{\PaperLanguage}[1]{}
\pacs{21.10.Ev, 21.60.Cs, 24.10.Pa}

\begin{document}
\begin{NoHyper}

\pdfoutput=1

%\title{Inclusion of the Longitudinal Component of the Momentum Transfer and Other Kinematic Factors in H(d,p)X Reactions}
%\title{Accounting for the Longitudinal Momentum-Transfer Component and Other Kinematic Factors in H(d,p)X Reactions}

\title{Inclusion of the Longitudinal Momentum-Transfer Component and
Kinematic Factors in a diffraction approach for H(d,p)X Reactions}

%\title{ ВРАХУВАННЯ ПОВЗДОВЖНЬОЇ КОМПОНЕНТИ ПЕРЕДАНОГО ІМПУЛЬСУ ТА ІНШИХ КІНЕМАТИЧНИХ ФАКТОРІВ У H(d,p)X }

\author{Yaroslav D. Krivenko-Emetov}

\affiliation{Institute for Nuclear Research, NAS of Ukraine, 03680, Kyiv}
\affiliation{National Technical University of Ukraine, 03056, Kyiv }

\email{krivemet@ukr.net, y.kryvenko-emetov@kpi.ua}

\author{Boris I. Sidorenko}
\affiliation{Institute for Nuclear Research, NAS of Ukraine, 03680, Kyiv }
\altaffiliation{Currently: pensioner. Affiliated with the Institute during 1994--2001. boris.sidorenko@meta.ua}

\begin{abstract}
%\abstract{%

In this work, within the framework of the Glauber–Sitenko approximation, an analysis of the differential cross section for deuteron breakup into a proton in the reaction $H(d,p)X$ is presented. The study is carried out using various parameterizations of the deuteron wave function, including the single-Gaussian parametrization, the multi-Gaussian K2 parametrization, and models based on the Av18 and NijmI nucleon–nucleon potentials.

Special attention is given to the effects of small longitudinal components of the transferred momentum ($Q_z<0.5$ GeV/c) and the transverse momentum of the proton–neutron pair ($|\mathbf{p}_\perp| <0.5 $ GeV/c) in the anti-laboratory reference frame. The results are compared with experimental data, particularly in the region of longitudinal momenta $p_3 = 0.25$–$0.5$~GeV/$c$, where quark effects are expected to manifest. Preliminary estimates show a decrease in the cross section with increasing transverse momentum, as well as a relatively small shift (and a significant increase) of the cross-section maximum due to the inclusion of the longitudinal component $Q_z$.

\end{abstract}
%\keywords{    дейтрон, дифракційний розпад, релятивістські ефекти, кваркова структура, переріз реакції, підхід Глаубера–Ситенка, антилабораторна система, повздовжній імпульс, поперечний імпульс }
\keywords{    deuteron,     diffractive breakup,     relativistic effects,     quark structure,     reaction cross section,     Glauber--Sitenko approach,     antilab frame,     longitudinal momentum,     transverse momentum,  dibarion, d*(2380)}

\maketitle

%====================================================================================
%7 добавив pdfplotsset для виправлення оберненої зумісності програми(bad log)
%\pgfplotsset{compat=1.18}
%\begin{document}
%\PaperLanguage{ukrainian} 
%\bigskip
%\ms
%\pacs{21.10.Ev, 21.60.Cs, 24.10.Pa}

\section{Introduction}
The deuteron, as the simplest bound nucleon system, remains a key subject of research in nuclear physics, particularly for understanding the non-central nature of nuclear forces, exchange and relativistic effects, and manifestations of the quark-gluon structure at short distances. Despite significant progress in the theoretical description of the deuteron, aspects such as its relativistic dynamics and behavior at high momentum transfers remain underexplored. These issues are closely linked to fundamental problems in the theory of strong interactions, particularly the application of quantum chromodynamics (QCD) to many-body problems (for details on these issues, see, e.g., \cite{Garcon}, \cite{Amarasinghe2023}, \cite{Kobushkin1998}, \cite{Brodsky2022}, \cite{Strikman2023}, \cite{Krivenko2024}).

One of the promising methods for studying the deuteron structure is the analysis of its breakup reactions on nuclei at high energies, particularly inclusive processes such as $A(d,p)X$. Experiments conducted at energies around 2.1–10.8 GeV/c~\cite{Ableev1983},\cite{Zaporozhets1986},\cite{Perdrisat1987},\cite{Ableev1988},\cite{Punjabi1989},\cite{Ableev1990},\cite{Nomofilov1994},\cite{Azhgirey1996},\cite{Cheung1992},\cite{Kuehn1994} revealed significant discrepancies between experimental data (including differential cross-sections, tensor analyzing power $T_{20}$, and polarization transfer) and predictions of traditional models based on the impulse approximation.

As noted in a number of studies, these discrepancies indicate the need to account for additional factors, such as the quark structure of the deuteron~\cite{Ableev1983},\cite{Kobushkin1998},\cite{Aono1995},\cite{Kobushkin1982},\cite{Azhgirey1996},\cite{Ierusalimov2010} and dilepton states~\cite{Sitenko2000},\cite{Platonova2023},\cite{Adlarson2014}.

Despite the success of the quark approach in describing the deuteron structure in hard processes, a number of open questions remain in $(d, p)$ breakup reactions. One of them concerns the role of Coulomb interaction between the target nucleus and the nucleons within the deuteron.

Experimental studies~\cite{Ableev1983,Ableev1987,Perdrisat1987,Ableev1988,%
Punjabi1989,Ableev1990},\cite{Nomofilov1994},\cite{Aono1995},%
\cite{Azhgirey1996},\cite{Cheung1992},\cite{Kuehn1994},%
\cite{AzhgireyJINR1996},\cite{Kuehn1994},%
\cite{Azhgirey1998} were conducted at small proton emission angles ($\theta_p < 7~\mu\mathrm{rad}$), corresponding to the conditions for applying the Glauber-Sitenko multiple-scattering diffraction theory~\cite{Glauber1959},\cite{Akhiezer1957},\cite{Nemetz}. Thus, this theory is applicable to studying these processes and predicts~\cite{Glauber1970} that Coulomb interaction may significantly affect the $A(d,p)X$-scattering cross-section.

Indeed, later, more detailed calculations within the Glauber-Sitenko model~\cite{Kobushkin2008} of deuteron breakup on carbon $C(d,p)X$ only improved the agreement between theory and experiment, showing that the influence of Coulomb interaction on the differential cross-section of $A(d, p)$ breakup is substantial only in the region of internal momenta $k < 50~$MeV. Consequently, this effect does not alter the conclusions regarding the importance of accounting for quark degrees of freedom in the deuteron at $k > 200~$MeV.

Moreover, Coulomb interaction, often neglected due to the dominance of the Coulomb barrier height, proves crucial for accurately describing the observed characteristics of deuteron breakup processes~\cite{Tartakovskyi2006},\cite{TartakovskyIvanova},\cite{Kobushkin2008},\cite{Davydovskyy2016}. However, as demonstrated in~\cite{Kobushkin1998},\cite{Krivenko2021}, in the studied reaction $H(d,p)X$, due to the significantly smaller nuclear charge compared to, for example, deuteron breakup on carbon $C(d,p)X$, its contribution is negligible.

However, in addition to the contribution of Coulomb interaction, several unresolved issues remain — particularly the inclusion of the longitudinal component of the transferred momentum, $Q_z$ (see, e.g.,~\cite{Davydovskyy2016}, \cite{Tartakovsky2005}, and references therein), which is neglected in classical diffraction theory~\cite{Glauber1959}, \cite{Akhiezer1957}, as well as the transverse components of the proton–neutron pair in the anti-laboratory reference frame (see, e.g.,~\cite{Sitnik2019}, \cite{Krivenko2021}).

Several studies~\cite{Davydovskyy2016}, \cite{Tartakovsky2005} demonstrate that neglecting these effects within the standard multiple-scattering diffraction theory (MSDT) can lead to misinterpretation of experimental results.

In this work, we present an analysis of the $H(d,p)X$ reaction, accounting for both the longitudinal transferred momentum component ($Q_z$) and the transverse momenta of nucleons in the anti-laboratory reference frame. Using various nucleon–nucleon interaction potentials, in particular the Nijmegen potential (Nijm I)~\cite{NijmI}, we study the impact of these factors on the differential cross section and compare the results with experimental data. Special attention is given to the region of high relative momenta ($k \geq 250$~МеВ/с), where traditional models show considerable discrepancies with the data~\cite{Ableev1983,Perdrisat1987,Ableev1988,Punjabi1989,Azhgirey1996}.
Thus, the aim of this study is to clarify the mechanism of deuteron breakup on nuclear targets by:
\begin{enumerate}[label=\arabic*.]
\item Including the longitudinal transferred momentum component $Q_z$ and the transverse momenta of nucleons in the anti-laboratory reference frame, and analyzing their impact on the cross-section shape.

\item Investigating how these effects influence the conclusions of previous studies concerning the manifestation of quark degrees of freedom in the short-range part of the deuteron wave function.
\end{enumerate}
The second objective warrants separate discussion. The relevance of this investigation is further underscored by the fact that soon after the pioneering studies~\cite{Ableev1983},\cite{Kobushkin1982}, where the observed rise in the invariant cross section in the anti-laboratory frame (in the range of 300-500 MeV/c) was attributed to the contribution of an $S$-wave quark component in the deuteron wave function, several alternative interpretations were proposed (e.g.,\cite{Braun1984},~\cite{Braun1986}), suggesting that the enhancement could result from pion rescattering. However, we argue that this interpretation raises certain doubts.

Indeed, the anti-laboratory frame in scattering problems is analogous to the $s$-channel: the location of maxima in the differential cross section can provide information about the mass of intermediate exchange particles, similarly to how resonance masses are described using the relativistic Breit–Wigner formula. If the aforementioned pions are on-shell, their energy would be considerably lower than the characteristic  300-500 MeV range.

At the same time, analysis of quark behavior in the nuclear medium shows that the effective masses of light quarks ($u$, $d$, $s$) increase significantly compared to their vacuum values, reaching 300-500 MeV ~\cite{Seely2009},\cite{Buballa2005},\cite{Saito2007},\cite{Bazavov2019},\cite{Aoki2017}. These data, at least qualitatively, contradict the assumption that pions (with masses of 135–140 MeV) play a dominant role in forming the observed enhancement.

Furthermore, a series of studies (see~\cite{Adlarson2014}) has shown that nucleon–nucleon collisions with total energy around 2380 MeV can give rise to a six-quark dibaryon resonance, $d^*(2380)$. Subtracting the deuteron mass (1876 MeV) yields an excess energy of about 500 MeV, consistent with the region where the enhancement is observed. Of particular interest is the work of the HAL QCD Collaboration~\cite{Gongyo2020}, which reports the first lattice QCD simulation directly addressing the possible existence of $d^*(2380)$.

That study investigates the $\Delta\Delta$ system with quantum numbers $I(J^P) = 0(3^+)$, one of the main candidates for the internal structure of the $d^*(2380)$ resonance, which was experimentally observed in $pn$-reactions at $\sqrt{s} \approx 2380$ MeV. Using the HAL QCD method, the authors constructed an effective $\Delta\Delta$ interaction potential for a range of pion masses (679–1018 MeV), and discovered a short-range (less than 1 fm), strongly attractive force that forms a quasi-bound state with a binding energy 25–40 MeV below the $\Delta\Delta$ threshold. These results qualitatively agree with the observed mass of $d^*(2380)$ and support its interpretation as a genuine six-quark resonance.

Although the arguments presented are mostly qualitative, they reinforce the hypothesis of a quark origin for the observed enhancement and confirm the relevance of further investigations. Importantly, the MSDT framework provides a natural way to include mesonic degrees of freedom — both explicitly, via realistic nucleon–nucleon potentials (such as Nijm I, which is based on meson exchange), and implicitly, through the parameters of profile functions. In any case, the observed enhancement lies within the kinematic domain where MSDT remains applicable, making it a promising independent tool for testing or refuting the quark-based interpretation.

The authors hope that the results of this work will be useful for future studies in relativistic nuclear physics, particularly in the context of interpreting experiments involving polarized deuteron beams.
\paragraph{Notation and frames (for clarity).}
Vectors are boldface (e.g., \(\mathbf{Q}\), \(\mathbf{p}\)); their magnitudes are plain symbols (e.g., \(Q \equiv |\mathbf{Q}|\), \(p \equiv |\mathbf{p}|\)). For any \(\mathbf{a}\), we write \(\mathbf{a}=(\mathbf{a}_\perp, a_z)\), with \(\mathbf{a}_\perp\!\cdot\!\hat{\mathbf{z}}=0\). All kinematic quantities are, unless stated otherwise, defined in the laboratory (LAB) frame where the incident deuteron moves along \(+\hat{\mathbf{z}}\). The “anti-laboratory frame’’ (ALF) denotes the frame where the deuteron is at rest (two-nucleon c.m. frame for the breakup). We use
$
\mathbf{p}_\perp \equiv \text{transverse component of the outgoing proton momentum in ALF and LAB}, ~
p_3 \equiv p_z \equiv \text{longitudinal proton component in LAB}.
$
The longitudinal transferred component is \(Q_z \equiv \mathbf{Q}\!\cdot\!\hat{\mathbf{z}}\) (LAB). Everywhere we compare or bound \(\mathbf{p}_\perp\) or \(\mathbf{Q}\) by scalars, we explicitly use their magnitudes, e.g. \(p_\perp \equiv |\mathbf{p}_\perp|\), \(Q \equiv |\mathbf{Q}|\). Units are consistently \(\mathrm{GeV}/c\) for momenta and \(\mathrm{fm}\) for lengths.

 \section{General Formalism}

As is well known (see~\cite{Glauber1959}, \cite{Akhiezer1957}, \cite{Nemetz}), the diffraction approximation is applicable only when the incident particle moves so rapidly that the target does not have time to significantly alter its configuration during the interaction. In nuclear processes, this allows one to simplify the description by neglecting the internal dynamics of nucleons in the target.

We also consider the case of scattering with small momentum transfer: the change in momentum $\mathbf{Q}$ of the incident particle is small compared to its initial momentum $\mathbf{p}$, that is
\[
|\mathbf{Q} |\ll |\mathbf{p}|.
\]
This assumption implies not only a small scattering angle but also a small transferred energy. Since high-energy particle scattering occurs predominantly at small angles, this case is of primary importance. In this regime, $\mathbf{Q}$ is small relative to $\mathbf{p}$, and thus the vector $\mathbf{Q}$ may be taken as orthogonal to the incident wave vector $\mathbf{p}$, i.e., lying in the $xy$-plane~\cite{LandauQM}.

Therefore, in the standard diffraction theory of interactions between composite particles and nuclei, it is assumed that each deuteron nucleon scatters off the target center with a phase shift determined solely by the transverse momentum transfer. This assumption is accurate for elastic scattering at small angles.

However, in the inelastic process of deuteron breakup, a nucleon may acquire not only transverse but also longitudinal momentum $Q_z$ (for example, when one of the deuteron nucleons is closer to the nucleus and is decelerated by nuclear forces, which is essentially the cause of the breakup). In this case, the longitudinal momentum effectively "tears" the deuteron apart. This effect is not included in the standard multiple-scattering diffraction theory (MSDT) and is not accounted for by the elastic profile function.

Nevertheless, it can be incorporated through a generalization of the MSDT by introducing small longitudinal momenta into the transition form factors and adding inelastic components to the profile function. As shown in several studies (see~\cite{Davydovskyy2016}, \cite{Tartakovsky2005}, \cite{Beiyusk}), inclusion of the longitudinal component can significantly affect the theoretical predictions.

We begin by considering the general Glauber--Sitenko formalism for the cross section of the deuteron breakup reaction $H(d, p)$, in the regime of longitudinal proton momentum $p_3 \sim p_d/2$ and small transverse momentum $\mathbf{p}_\perp$ (The deuteron momentum is given as \(\mathbf{p}_d = (0, 0, p_d)\). In this case, as previously mentioned, inelastic scattering appears as a natural modification of the elastic term.

Although the Coulomb interaction is important for accurately reproducing the experimental characteristics of the breakup process, in the specific reaction of deuteron breakup on a proton, the Coulomb interaction effectively involves only the two protons (neglecting the neutron's electric form factor). Therefore, to first approximation, this interaction may be neglected.

Neglecting the Coulomb interaction, the strong interaction amplitude of deuteron breakup $F_{\text{str.}}(\mathbf{p}_\perp, p_3, \mathbf{Q})$ takes the form:
\begin{equation}
F_{\text{str.}}(\mathbf{p}_\perp, p_3, \mathbf{Q}) = \frac{i p_d}{2\pi} \int d^2\mathbf{B} \, e^{i \mathbf{Q}_\perp \cdot \mathbf{B}} \int d^3\mathbf{r} \, \psi_{\mathbf{k}}^{*(-)}(\mathbf{r}) (1-S(\mathbf{B}, \mathbf{r})) \psi_0(\mathbf{r}) ,
\label{eq:Fstr}
\end{equation}
where $ \psi_0(\mathbf{r})$ is the deuteron wave function, $\psi_{\mathbf{k}}^{*(-)}(\mathbf{r})$ is the distorted final-state wave function of the outgoing $pn$-pair, ${\mathbf{k}}$ is the wave number of the incoming deuteron and $S(\mathbf{B}, \mathbf{r})$ is the Glauber--Sitenko scattering operator that includes the effect of longitudinal and transverse momenta via impact parameter $\mathbf{B}$ and internal coordinate $\mathbf{r}$.

The operator $S(\mathbf{B}, \mathbf{r})$ describes the total scattering effect of both deuteron nucleons interacting with the target, taking into account the phase shifts acquired while passing through the nuclear field. Within the optical approximation, it is given by
\[
S(\mathbf{B}, \mathbf{r}) = 1 - \Gamma_1(\mathbf{B} + \tfrac{1}{2} \mathbf{r_\perp}) - \Gamma_2(\mathbf{B} - \tfrac{1}{2} \mathbf{r_\perp}) + \Gamma_1(\mathbf{B} + \tfrac{1}{2} \mathbf{r_\perp}) \Gamma_2(\mathbf{B} - \tfrac{1}{2} \mathbf{r_\perp}),
\]
where $\mathbf{r_\perp}$ is the projection of the vector $\mathbf{r}$ onto the plane perpendicular to the scattering direction (the $xy$ plane), and $\Gamma_1$, $\Gamma_2$ are the profile functions of the deuteron nucleons. In a simplified approximation, the $\Gamma_i$ can be expressed in terms of nucleon-nucleon phase shifts.

The inclusion of the longitudinal momentum component $Q_z$ affects both the final-state wave function and the shape of the profile function, particularly in the case of inelastic scattering. This effect can be incorporated through the dependence of $S(\mathbf{B}, \mathbf{r}, Q_z) =e^{i \chi_{\text{str.}}(\mathbf{B},\mathbf{r}, Q_z)}$ on the full coordinate $\mathbf{r}$, especially its $z$-component.

Further specification of the analytical form of $F_{\text{str.}}$ requires choosing specific models for  the profile function and the wave functions, which in turn depends on the nucleon-nucleon interaction potential (e.g., the Nijm~I potential), as well as the approximation used for the final-state $pn$ pair.

Introducing ~$\mathbf{B} = \mathbf{b}_p - \frac{\mathbf{r}_\perp}{2}$ and ~$\mathbf{B} = \mathbf{b}_n + \frac{\mathbf{r}_\perp}{2}$, the formula (\ref{eq:Fstr}) can be rewritten as follows:

\begin{align}
F_{\text{str.}}(\mathbf{p}_\perp, p_3, \mathbf{Q}) ={} & \frac{i p_d}{2 \pi} \int d^2 \mathbf{B} \int d^3 \mathbf{r}  e^{i \mathbf{Q}_\perp\cdot \mathbf{B}} \,\psi^*_k(\mathbf{r}) \, \psi_0(\mathbf{r})  \left( 1 - e^{i \chi_{\text{str.}}(\mathbf{b}_p, \mathbf{b}_n, Q_z)} \right) \nonumber \\
= {} & \frac{i p_d}{2 \pi} \int d^2  \mathbf{B}  \int d^3  \mathbf{r} e^{i  \mathbf{Q}_\perp \cdot  \mathbf{B} } \,\psi^*_k( \mathbf{r}) \, \psi_0( \mathbf{r}) \left( \Gamma( \mathbf{b}_n, Q_z) + \Gamma( \mathbf{b}_p, Q_z) - \Gamma( \mathbf{b}_n, Q_z) \Gamma( \mathbf{b}_p, Q_z) \right) \nonumber \\
= {} & F_{\text{str.}}^n (\mathbf{p}_\perp, p_3, \mathbf{Q}) + F_{\text{str.}}^p(\mathbf{p}_\perp, p_3, \mathbf{Q}) - F_{\text{str.}}^{np} (\mathbf{p}_\perp, p_3, \mathbf{Q})
\label{eq:Fstr1}
\end{align}
Here:

\begin{itemize}
  \item \( \mathbf{p}_\perp \) — transverse component of the proton momentum in the laboratory frame;
  \item \( p_3 \) — longitudinal component of this momentum;
  \item \( \mathbf{Q} = (\mathbf{Q}_\perp, Q_z) \) — momentum transferred to the proton-neutron system;
  \item \( \mathbf{B} = \frac{1}{2} (\mathbf{b}_p + \mathbf{b}_n) \) — average transverse impact parameter (impact center);
  \item \( \mathbf{r}_\parallel = r_z\) — longitudinal component of the vector \( \mathbf{r} = \mathbf{r}_p - \mathbf{r}_n \);

  \item \( \chi_{\text{str.}}(\mathbf{b}_p, \mathbf{b}_n, Q_z) \) — strong scattering phase;
  \item \( \Gamma(\mathbf{b}_p, Q_z), \Gamma(\mathbf{b}_n, Q_z) \) — profile functions describing the interaction of the proton and neutron with the target;
  \item \( F_{\text{str.}}^n, F_{\text{str.}}^p, F_{\text{str.}}^{np} \) — structural amplitudes corresponding to the contributions from neutron scattering, proton scattering, and their interference.
\end{itemize}

This approach allows one to isolate the contributions from each nucleon, as well as their interaction in diffractive scattering.

The quantities \(\mathbf{B} = \frac{1}{2} (\mathbf{b}_p + \mathbf{b}_n)\) and \(\mathbf{r}_\perp = \mathbf{b}_p - \mathbf{b}_n\) are defined in terms of the impact parameters \(\mathbf{b}_p\) (proton) and \(\mathbf{b}_n\) (neutron). 

The definition of the relative momentum between the proton and neutron, \( \mathbf{k} \), requires additional clarification:

\begin{itemize}
    \item In the \textbf{non-relativistic case}, it is expressed as:
    \[
    \mathbf{k} = \left( \mathbf{p}_\perp - \frac{1}{2} \mathbf{Q}_\perp,\, p_3 \right).
    \]
    
    \item For a \textbf{relativistic deuteron} (\( p_d \gg m_d \)), it is defined by transitioning to a reference frame in which the total longitudinal momentum of the two-nucleon system is zero:
    \[
    p_3^* + n_3^* = 0, \quad \mathbf{p}_\perp^* = \mathbf{p}_\perp, \quad \mathbf{n}_\perp^* = \mathbf{n}_\perp.
    \]
    The components of the relative momentum then take the form:
    \[
    \mathbf{k}_\perp = \frac{1}{2} (\mathbf{p}_\perp - \mathbf{n}_\perp) = \mathbf{p}_\perp - \frac{1}{2} \mathbf{Q}_\perp, \quad 
    k_3 = \frac{1}{2} (p_3^* - n_3^*) = p_3^*,
    \]
    where \( \mathbf{p}^* \) and \( \mathbf{n}^* \) are the proton and neutron momenta in the so-called anti-laboratory frame, while \( \mathbf{p} \) and \( \mathbf{n} \) refer to their momenta in the laboratory frame. This definition is valid in the vicinity of \( p_3 \sim \frac{1}{2} p_d \).
\end{itemize}

In Ref.~\cite{Kobushkin2008}, the perpendicular component of the relative momentum in the anti-laboratory frame, $\mathbf{p}_\perp$, was set to zero. However, in reality, due to the uncertainty principle, at distances comparable to the deuteron size, the momentum uncertainty should be of the order of the meson mass (130--150~MeV). Experimental measurements yield average values close to this scale.

Applying the Lorentz transformation
\[
p_3 = \gamma p_3^* + \gamma \beta \sqrt{(p_3^*)^2 + M^2 + p_\perp^2} = \frac{1}{2} p_d + e_p,
\]
where $e_p$ is the excess momentum reflecting the asymmetry in the distribution of $p_d$ between the proton and neutron, from the laboratory frame (in which the deuteron moves with longitudinal momentum $p_3$) to the anti-laboratory frame (in which it is at rest), and assuming the proton and neutron momenta in the deuteron rest frame are $p^*$ and $n^*$, respectively, we obtain the following simplified expression for small $p^*$ (see Appendix~A):

\begin{align*}
p_3 &\approx \underbrace{\frac{p_d}{M_d} \sqrt{M^2 + p_\perp^2}}_{\text{leading term}} 
+ \underbrace{\frac{E_d}{M_d} p_3^*}_{\text{linear correction}} 
+ \underbrace{\frac{p_d}{2M_d} \frac{(p_3^*)^2}{\sqrt{M^2 + p_\perp^2}}}_{\text{quadratic correction}} \\
&= \frac{1}{M_d} \left( p_d \sqrt{M^2 + p_\perp^2} + E_d p_3^* + \frac{p_d (p_3^*)^2}{2\sqrt{M^2 + p_\perp^2}} \right),
\end{align*}
where $M$ is the nucleon mass, $M_d$ is the deuteron mass,  $E_d$ is the energy in the laboratory frame and $p_3^*$ is the longitudinal component of the nucleon momentum in the deuteron rest frame.

We observe that when \( p_3^* = 0 \) and \( p_\perp = 0 \), the proton’s longitudinal momentum—consistent with physical expectations—equals its mean value \( p_3 = p_d / 2 \).  

However, since our focus is on the relationship between the proton-neutron pair’s relative momentum in the anti-laboratory frame and the laboratory frame (specifically, its deviation from \( p_d / 2 \) in the latter), we obtain:  

\[
d p_3 = e_p =p_3-\frac{1}{2} p_d  \approx  \frac{1}{M_d} \left( E_d p_3^* + \frac{p_d (p_3^*)^2}{2\sqrt{M^2 + p_\perp^2}} \right)
\]

Expressing the impact parameter in terms of the transverse component of the relative position vector of the proton–neutron pair (~ $
\mathbf{B} = \mathbf{b}_p - \frac{\mathbf{r}_\perp}{2} \quad \text{and} \quad \mathbf{B} = \mathbf{b}_n + \frac{\mathbf{r}_\perp}{2}$),
and substituting these into the amplitude expression, we obtain:

\begin{align}
F_{\text{str.}}(\mathbf{p}_\perp, p_3, \mathbf{Q}) 
={} & \, \frac{i p_d}{2 \pi} \int d^2 \mathbf{b}_n \, e^{i  \mathbf{Q}_\perp \cdot  \mathbf{b}_n}\Gamma( \mathbf{b}_n)  
\int d^3  \mathbf{r} \, e^{- i \mathbf{Q} \cdot \mathbf{r}/2} \,
\psi^*_k(\mathbf{r}) \, \psi_0(\mathbf{r}) \nonumber \\
& + \frac{i p_d}{2 \pi} \int d^2 \mathbf{b}_p \, e^{i  \mathbf{Q}_\perp \cdot  \mathbf{b}_p} \Gamma( \mathbf{b}_p)
\int d^3  \mathbf{r} \, e^{ i \mathbf{Q} \cdot \mathbf{r}/2} \,
\psi^*_k(\mathbf{r}) \, \psi_0(\mathbf{r}) \nonumber \\
& - \frac{i p_d}{2 \pi}  \int d^2  \mathbf{B}  \int d^3  \mathbf{r} e^{i  \mathbf{Q}_\perp \cdot  \mathbf{B} } \,\psi^*_k( \mathbf{r}) \, \psi_0( \mathbf{r}) \times \Gamma( \mathbf{b}_p) \Gamma( \mathbf{b}_n) 
\label{Fstr1}        
\end{align}

Here we have selected the inelastic scattering profile functions accounting for the longitudinal momentum component $Q_z$ in the most general form, including both linear and quadratic terms in the exponential argument (see \cite{Davydovskyy2016} for dependencies in the phase $~iQ_z/2$, and \cite{Tartakovsky2005} with \cite{Beiyusk} for $~Q_z^2$ behavior), as follows:

 \begin{equation}
 \Gamma( \mathbf{b}_p, Q_z) =\Gamma( \mathbf{b}_p) e^{ i Q_z \cdot r_{\parallel}/2} = \frac{ e^{ - i \mathbf{Q}_z \cdot \mathbf{r}_{\parallel}/2} }{2 \pi i p_p} \int d^2 \mathbf{l }\, e^{-i\mathbf{l}\mathbf{b_p}} f_p(\mathbf{l}) 
    =   \frac{(1- i \rho_p) \sigma_p}{4 \pi \beta_p^2} 
e^ {- \frac{i}{2} \mathbf{Q}_z \cdot \mathbf{r}_{\parallel} - \frac{1}{2}\left(\frac{\mathbf{b}_p^2}{\beta_p^2} +\alpha \beta_p^2 Q_z^2 \right)} ,
\end{equation}
and
\begin{equation}
 \Gamma( \mathbf{b}_n, Q_z) = \Gamma( \mathbf{b}_n)  e^{- i Q_z \cdot r_{\parallel}/2}= \frac{ e^{ i \mathbf{Q}_z \cdot \mathbf{r}_{\parallel}/2} }{2 \pi i p_n} \int d^2\mathbf{ l} \, e^{-i\mathbf{l}\mathbf{b_n}} f_n(\mathbf{l})   
  = \frac{(1- i \rho_n) \sigma_n}{4 \pi \beta_n^2} 
e^{ \frac{i}{2} \mathbf{Q}_z \cdot \mathbf{r}_{\parallel} 
- \frac{1}{2}\left(\frac{\mathbf{b}_n^2}{\beta_n^2} +\alpha \beta_n^2 Q_z^2 \right) } ,
\end{equation}

where the amplitude of nucleon scattering on the nucleus takes the form:

\[
f_N(\mathbf{l}) = \frac{(i + \rho_N)\, p_N\, \sigma_N}{4\pi} \exp\left[-\frac{1}{2} \beta_N^2 \left( \mathbf{l}^2 + \alpha Q_z^2 \right)\right], \quad (N = n, p),
\]

where \( \sigma_N \) is the total nucleon-proton scattering cross section,
and \( \rho_N \) is the ratio of the real to imaginary parts of the amplitude.

According to the optical theorem, at the considered energies the \( \rho_N \) parameter is small, while the cross sections, following Pomeranchuk's theorem, differ only slightly.

Therefore, if - as both theory and experiment suggest - \( \sigma_p \approx \sigma_n \), we will subsequently use the following natural approximation:
\[ \sigma_p \simeq \sigma_n \equiv \sigma \quad \text{and} \quad \rho_N \ll 1. \]

Here, $\alpha$ is a parameter that relates $Q_z$ to the deuteron's internal structure (e.g., through its momentum distribution). For the deuteron, $\alpha \sim \mathcal{O}(1)$, given its $\sim 1$~fm size, while $\beta_N$ (known as the nuclear slope parameter for nucleon-nucleus scattering) corresponds to the range of the nucleon-nucleon interaction ($\sim 1$~fm).

These parameters determine the physical significance of both terms in the exponential:

\begin{itemize}
  \item $\alpha \beta_N^2 Q_z^2$ -- governs the deuteron's $Q_z$-directional breakup (resulting from the differential longitudinal momentum transfer to the proton and neutron during scattering, which disrupts/stretches the deuteron)
  \item $b_N^2 / \beta_N^2$ -- controls transverse spatial damping (where the Gaussian width in momentum space sets the nuclear spatial scale)
\end{itemize}
Therefore, we can neglect the $\alpha \beta_N^2 Q_z^2$ term when it is significantly smaller than the dominant exponential damping term, i.e.:
\[
\alpha \beta_N^2 Q_z^2 \ll \frac{b_N^2}{\beta_N^2} 
\quad \Rightarrow \quad 
Q_z^2 \ll \frac{b_N^2}{\alpha \beta_N^4}
\]

Alternatively stated: the longitudinal momentum transfer $Q_z$ is small when:
\[
|Q_z| \ll \frac{|b_N|}{\beta_N^2 \sqrt{\alpha}}
\]

Since $1/\beta_N$ determines the profile width in momentum space (i.e., the spreading radius of $f_N(\mathbf{l})$), we have:
\[
1/\beta_N \sim \frac{1}{\text{nuclear size}} \approx \frac{1}{1~\text{fm}} \approx 0.2~\text{GeV}
\quad \text{and} \quad
|b_N| \sim 1 - 3~\text{fm}
\]

This yields the approximate estimate:
\[
|Q_z| \ll \frac{1-3~\text{fm}}{(1~\text{fm})^2 \sqrt{\alpha}} 
= \frac{5-15~\text{GeV}^{-1}}{25~\text{GeV}^{-2} \sqrt{\alpha}} 
\approx \frac{0.2-0.6}{\sqrt{\alpha}}~\text{GeV}
\]
%\subsection*{Dimensional meaning and estimate of the parameter $\alpha$}

In our profile function the exponential has the form
\begin{equation}
  \exp\!\left[-\frac{1}{2}\,\beta_N^2\!\left(l^{2}+\alpha\,Q_z^{2}\right)\right].
  \label{eq:exp-alpha}
\end{equation}
Here $l$ denotes a transverse relative momentum and $Q_z$ is the longitudinal
momentum transfer; $\beta_N$ is a momentum--scale parameter (with dimension
$[\beta_N]=\text{GeV}^{-1}$ in natural units $\hbar=c=1$).

Since the combination $\beta_N^2 l^2$ in~\eqref{eq:exp-alpha} is dimensionless,
the factor multiplying $Q_z^2$ must also be dimensionless. Therefore,
\[
  \alpha \ \text{is dimensionless.}
\]
Physically, $\alpha$ controls the relative ``width'' (i.e., the typical scale)
of the longitudinal structure compared to the transverse one that is set by
$\beta_N$.

\paragraph{Characteristic length/momentum scales.}
Let $b_N$ be the transverse spatial scale of the nucleon--nucleon interaction
(the Gaussian width in coordinate space). Then, in natural units,
\[
  \beta_N \sim \frac{1}{b_N}.
\]
For nuclear sizes we take
\[
  b_N \sim 1\text{--}3~\mathrm{fm}
  \qquad\Longrightarrow\qquad
  \beta_N \sim \frac{1}{b_N} \sim 3\text{--}5~\mathrm{GeV}^{-1}.
\]
For the deuteron, a typical longitudinal size is
\[
  L_z \sim 1\text{--}2~\mathrm{fm}.
\]

\paragraph{Natural estimate for $\alpha$.}
A convenient and physically transparent estimate is to relate $\alpha$ to the
ratio of the longitudinal to transverse sizes:
\begin{equation}
 \ \alpha \ \sim \ \left(\frac{L_z}{\,b_N\,}\right)^{\!2}\ .
  \label{eq:alpha-estimate}
\end{equation}
Using $L_z \sim 1\text{--}2~\mathrm{fm}$ and $b_N \sim 1\text{--}3~\mathrm{fm}$
gives the (order-of-magnitude) range
\begin{equation}
  0.3 \ \lesssim\  \alpha \ \lesssim\ 4,
\end{equation}
with the upper end ($\alpha\!\sim\!4$) corresponding to a more extended
longitudinal structure ($L_z\!\approx\!2~\mathrm{fm}$, $b_N\!\approx\!1~\mathrm{fm}$),
and the lower end ($\alpha\!\sim\!0.3$) to a relatively broader transverse scale.
Thus, for typical nuclear/deuteron dimensions, one expects $\alpha$ to be a
quantity of order unity:
\[
  \alpha=\mathcal{O}(1).
\]

Thus, for $Q_z \lesssim 1~\text{GeV}$ and $\alpha \sim 1$, the $\alpha \beta_N^2 Q_z^2$ term can be safely neglected.

\bigskip

This is equivalent to stating that the longitudinal momentum transfer $Q_z$ is substantially smaller than the characteristic transverse structure scales set by $\Gamma(b_N)$. In practice, for experiments with $Q_z \lesssim 1$~GeV/c, the $Q_z^2$ term can be neglected. However, in this work we conservatively adopt $Q_z \lesssim 0.5$~GeV/c. Naturally, when studying more subtle effects or at higher longitudinal momenta (larger angles), one must account for the $Q_z^2$-dependence (see, e.g., \cite{Tartakovsky2005}).

With these considerations, the profile functions in our approximation take the form:

 \begin{equation}
 \Gamma( \mathbf{b}_p, Q_z) \sim   \frac{(1- i \rho_p) \sigma_p}{4 \pi \beta_p^2} 
\exp\left(- \frac{i}{2} \mathbf{Q}_z \cdot \mathbf{r}_{\parallel} 
- \frac{1}{2}\left(\frac{\mathbf{b}_p^2}{\beta_p^2}  \right) 
\right) ,
\end{equation}
and
\begin{equation}
 \Gamma( \mathbf{b}_n, Q_z)  \sim \frac{(1- i \rho_n) \sigma_n}{4 \pi \beta_n^2} 
\exp\left( \frac{i}{2} \mathbf{Q}_z \cdot \mathbf{r}_{\parallel} 
- \frac{1}{2}\left(\frac{\mathbf{b}_n^2}{\beta_n^2}  \right) 
\right) ,
\end{equation}
Then, introducing the standard definition for the transferred form factor:
\begin{equation}
  G\left(-\tfrac{1}{2} \mathbf{Q}, \mathbf{k} \right)   =  \int d^3  \mathbf{r} \, e^{ i \mathbf{Q} \cdot \mathbf{r}/2} \psi^*_k(\mathbf{r}) \, \psi_0(\mathbf{r}).
\end{equation}
The proton amplitude in (\ref{Fstr1}) can be rewritten as follows:

\[
F_{\text{str.}}^{p} (\mathbf{p}_\perp, p_3, \mathbf{Q}) = G\left(-\tfrac{1}{2} \mathbf{Q}, \mathbf{k} \right) \cdot \frac{i p_d}{2\pi} \cdot \frac{1}{2\pi i p_p} \int d^2 \mathbf{b}_p \int d^2 \mathbf{l}_p \, e^{i \mathbf{b}_p \cdot (\mathbf{Q}_\perp - \mathbf{l}_p)} f_p(\mathbf{l}_p).
\]

The integral over $ \mathbf{b}_p $ yields a delta function:

\[
\int d^2 \mathbf{b}_p \, e^{i \mathbf{b}_p \cdot (\mathbf{Q}_\perp - \mathbf{l}_p)} = (2\pi)^2 \delta^{(2)}(\mathbf{Q}_\perp - \mathbf{l}_p).
\]
Substituting the delta-function into the amplitude expression and simplifying, obtain:
\[
F_{\text{str.}}^{p} (\mathbf{p}_\perp, p_3, \mathbf{Q}) = G\left(-\tfrac{1}{2} \mathbf{Q}, \mathbf{k} \right) \cdot \frac{p_d}{p_p} \int d^2 \mathbf{l}_p \, \delta^{(2)}(\mathbf{Q}_\perp - \mathbf{l}_p) f_p(\mathbf{l}_p) = G\left(-\tfrac{1}{2} \mathbf{Q}, \mathbf{k} \right) \cdot \frac{p_d}{p_p} f_p(\mathbf{Q}_\perp).
\]

Next, using in the amplitude the expression $ f_p(\mathbf{Q}_\perp) = \frac{(i + \rho_p)\, p_p\, \sigma_p}{4\pi} \exp\left( -\tfrac{1}{2} \beta_p^2 \mathbf{Q}_\perp^2 \right)$

after straightforward simplifications, obtain:

\[
F_{\text{str.}}^{p} (\mathbf{p}_\perp, p_3, \mathbf{Q}) = \frac{(i + \rho_p)\, \sigma_p\, p_d}{4\pi} \cdot G\left(-\tfrac{1}{2} \mathbf{Q}, \mathbf{k} \right) \exp\left( -\tfrac{1}{2} \beta_p^2 \mathbf{Q}_\perp^2 \right).
\]

Following similar transformations, the neutron amplitude expression becomes:
\[
F_{\text{str.}}^{n} (\mathbf{p}_\perp, p_3, \mathbf{Q}) = \frac{(i + \rho_n)\, \sigma_n\, p_d}{4\pi} \, G\left(\tfrac{1}{2} \mathbf{Q}, \mathbf{k}\right) \exp\left(-\tfrac{1}{2} \beta_n^2\, \mathbf{Q}_\perp^2\right)
\]

% Підставлення виразу для \Gamma(b) у F_{pn}
%\paragraph{Підстановка профільної функції та перетворення координат}
%Тепер розглянемо у (\ref{Fstr1}) вираз для амплітуди протон-нейтронного перерозсіяння \( F_{pn} \). Починаємо з амплітуди:

Substituting into this expression the explicit forms of \( \Gamma_p \) and \( \Gamma_n \), namely,
\[
\Gamma_{p,n}(\mathbf{b}) = \frac{1}{2\pi i p_{p,n}} \int d^2 \mathbf{l}_{p,n} \, e^{-i \mathbf{l}_{p,n} \cdot \mathbf{b}} f_{p,n}(\mathbf{l}_{p,n}),
\]
we obtain an expression for \( F_{pn} \) in terms of the nucleon amplitudes \( f_p(\mathbf{l}_p) \) and \( f_n(\mathbf{l}_n) \):

\[
F_{pn} = \frac{i p_d}{2 \pi} \int d^2 \mathbf{B} \int d^3 \mathbf{r} \, \psi_k(\mathbf{r}) e^{i \mathbf{B} \cdot \mathbf{Q}_\perp} \psi_0(\mathbf{r}) \left( \frac{1}{2 \pi i p_p} \int d^2 \mathbf{l}_p e^{-i \mathbf{l}_p \cdot \mathbf{b}_p} f_p(\mathbf{l}_p) \right) \left( \frac{1}{2 \pi i p_n} \int d^2 \mathbf{l}_n e^{-i \mathbf{l}_n \cdot \mathbf{b}_n} f_n(\mathbf{l}_n) \right),
\]
%де \(  \mathbf{B}= \mathbf{b}_p - \frac{\mathbf{r}_\perp}{2} \), \( \mathbf{b}_n = \mathbf{b}_p - \mathbf{r}_\perp \).

% Переписуємо F_{pn} з урахуванням G

Next, using the definition of the transition form factor ~$ G\left(-\frac{1}{2} \mathbf{Q}_\perp + \mathbf{l}_n, \mathbf{k}\right) = \int d^3 \mathbf{r} e^{-i \frac{1}{2} \mathbf{Q}_\perp \cdot \mathbf{r}_\perp + i \mathbf{l}_n \cdot \mathbf{r}_\perp} \psi_k^*(\mathbf{r}) \psi_0(\mathbf{r}) $, and performing the substitutions $\mathbf{B} = \mathbf{b}_p - \frac{\mathbf{r}_\perp}{2}$ and $\mathbf{b}_n = \mathbf{b}_p - \mathbf{r}_\perp$, we obtain:

\[
F_{\text{str.}}^{pn} = \frac{i p_d}{2 \pi} \int d^2 \mathbf{l}_p f_p(\mathbf{l}_p) \int d^2 \mathbf{l}_n G\left(-\frac{1}{2} \mathbf{Q}_\perp + \mathbf{l}_n, \mathbf{k}\right) f_n(\mathbf{l}_n) \int d^2 \mathbf{b}_p e^{i \mathbf{b}_p \cdot \mathbf{Q}_\perp} \left( \frac{1}{2 \pi i p_p} e^{-i \mathbf{l}_p \cdot \mathbf{b}_p} \right) \left( \frac{1}{2 \pi i p_n} e^{-i \mathbf{l}_n \cdot \mathbf{b}_p} \right).
\]

% Об'єднання експонент
 
After combining the exponentials: $ e^{i \mathbf{b}_p \cdot \mathbf{Q}_\perp} e^{-i \mathbf{l}_p \cdot \mathbf{b}_p} e^{-i \mathbf{l}_n \cdot \mathbf{b}_p} = e^{i \mathbf{b}_p \cdot (\mathbf{Q}_\perp - \mathbf{l}_p - \mathbf{l}_n)} $, the integral over $ \mathbf{b}_p $ yields a two-dimensional delta function:

\[
\int d^2 \mathbf{b}_p e^{i \mathbf{b}_p \cdot (\mathbf{Q}_\perp - \mathbf{l}_p - \mathbf{l}_n)} = (2\pi)^2 \delta^{(2)}(\mathbf{Q}_\perp - \mathbf{l}_p - \mathbf{l}_n).
\]

% Підставлення дельта-функції
Substituting this delta function into the expression for the rescattering amplitude, after straightforward simplifications, we finally obtain:
\[
F_{\text{str.}}^{pn} (\mathbf{p}_\perp, p_3, \mathbf{Q}) = -\frac{i p_d}{2 \pi p_p p_n} \int d^2 \mathbf{l}_p \, f_p(\mathbf{l}_p) \, G\left(\frac{1}{2} \mathbf{Q}_\perp - \mathbf{l}_p, \mathbf{k}\right) f_n(\mathbf{Q}_\perp - \mathbf{l}_p).
\]

\section{Wave Functions and Normalization Issues}

To draw more general conclusions, both analytical and numerical calculations were carried out for a broad class of deuteron wave functions of Gaussian type. Specifically, the calculations were performed using the single-Gaussian parametrization by Tartakovsky~\cite{Tartakovsky2005}, the multi-Gaussian parametrization K2~\cite{Simenog} as well as the realistic AV18~\cite{Wiringa1995} and Nijm-I~\cite{NijmI} potentials.

As an example, let us consider in more detail the normalization condition for the Nijm-I potential (for other Gaussian-type potentials the conclusions will be similar). In this case, the $S$- and $D$-wave components of the wave function are approximated by a sum of Gaussians:

$$u(r) = r \sum_{i=1}^N A_i e^{-\alpha_i r^2}, \quad w(r) = r^3 \sum_{i=1}^N B_i e^{-\beta_i r^2}.$$

As is well known, the bound and normalized state of the deuteron has the following form:

\be\label{eq:21}
\psi_{\text{bound}}(\mathbf{r}) =\psi_{0}(\mathbf{r})= \sqrt{N_s} \, \psi_s(\mathbf{r}) + \sqrt{N_d} \, \psi_d(\mathbf{r}) = \frac{1}{(4\pi)^{1/2} r} ( u(r) +w(r)) ,
\ee
%де $N_s$ й $N_d$ --- імовірністі $S$ та $D$ -хвилі  в дейтроні \( (|N_s| + |N_d| = 1) \), а \(\psi_s(\mathbf{r})\) і \(\psi_d(\mathbf{r})\) — нормалізовані та ортогональні хвильові функції відповідних S й D - станів:

where \( N_s \) and \( N_d \) are the probabilities of the \( S \)- and \( D \)-wave components in the deuteron \( (|N_s| + |N_d| = 1) \), and \( \psi_s(\mathbf{r}) \) and \( \psi_d(\mathbf{r}) \) are the normalized and orthogonal wave functions of the corresponding \( S \)- and \( D \)-states:

\be\label{eq:22}
\int \psi_s^*(\mathbf{r}) \psi_s(\mathbf{r}) \, d^3r = 1, \quad \int \psi_d^*(\mathbf{r}) \psi_d(\mathbf{r}) \, d^3r = 1, \quad \int \psi_s^*(\mathbf{r}) \psi_d(\mathbf{r}) \, d^3r = 0.
\ee

As is known, for the Gaussian wave function (\ref{eq:21}), it is impossible to construct a wave function for the unbound $pn$ system that would strictly satisfy the conditions of orthogonality and completeness (see, for example, \cite{Evlanov1980}). Therefore, following the approach of \cite{Evlanov1980}, we will first construct only a function orthogonal to the deuteron wave function. 

In the general case, following \cite{Kobushkin2008}, we write the unbound deuteron state in the form:

\be\label{eq:23}
\psi_{\mathbf{k}}(\mathbf{r}) = D e^{i \mathbf{k} \cdot \mathbf{r}} - C \left( \tilde{\varphi}_s(\mathbf{k}) \psi_s(\mathbf{r}) + \tilde{\varphi}_d(\mathbf{k}) \psi_d(\mathbf{r}) \right) \sim D e^{i \mathbf{k} \cdot \mathbf{r}} - C \left( \tilde{\varphi}_s(\mathbf{k}) \psi_s(\mathbf{r}) \right) .
\ee

where \( \psi_s(r) = \frac{1}{(4\pi)^{1/2} r} \, \frac{u(r)}{\sqrt{N_s}} \), and \( \tilde{\varphi}_s(\mathbf{k}) \) and \( \tilde{\varphi}_d(\mathbf{k}) \) are the complex conjugate Fourier transforms of the \( S \)- and \( D \)-components of the deuteron wave function. 

For example, for the single-Gaussian parametrization of the \( S \)-wave~\cite{Tartakovsky2005}, we have:

\begin{equation}
 \tilde{\varphi}_s(\mathbf{k}) = \frac{1}{(2\pi)^{3/2}}\int \psi_s^*(\mathbf{r}) e^{i\mathbf{k}\cdot\mathbf{r}} d^3\mathbf{r}= \tilde{\phi}_s^*(\mathbf{k})= \sqrt{\frac{1}{4\pi}} \sum_{i=1}^{N} \frac{A_i}{(2\alpha_i)^{3/2}} e^{-k^2/(4\alpha_i)}
\end{equation}

The fulfillment of the orthogonality condition  
\[
\int \psi_{\mathbf{k}}^*(\mathbf{r}) \, \psi_{\text{bound}}(\mathbf{r}) \, d^3r = 0
\]
requires the following relation between the normalization constants (see Appendix~B and \cite{Kobushkin2008}):
\[
C = (2\pi)^{3/2} D.
\]

However, under this condition, it becomes impossible to simultaneously satisfy the normalization requirement for the continuous spectrum (i.e., the completeness condition), which demands the delta-function normalization:

\begin{equation}
\langle \psi_{\mathbf{k}'} | \psi_{\mathbf{k}} \rangle = (2\pi)^{3}\delta^{(3)}(\mathbf{k} - \mathbf{k}')
\end{equation}

This condition is satisfied only approximately when $\alpha_i \to 0$, and then, for instance, for the single-Gaussian parametrization \cite{Tartakovsky2005}, $D \to 1$ at non-zero $\mathbf{k} - \mathbf{k}'$ (see Appendix $\mathbf{CI}$). Similar conditions for the constants $D$ and $C$ are obtained in the approximate normalization of the continuous spectrum at zero $\mathbf{k} - \mathbf{k}'$ (see Appendix $\mathbf{CII}$).

\section{Comparison with Experiment}

Following the works~\cite{DeForest1966, Donnelly1975}, the wave function can be expressed in terms of conventional two-nucleon deuteron wave functions, renormalized according to the condition formulated in~\cite{Kobushkin1978}, \cite{Kobushkin2008}.

The differential invariant cross section is defined by the formula

\begin{equation}
\label{eq:4.24}
E_p\frac{d^3\sigma}{d^3k} =E_p \frac{1}{(2\pi)^3} \int d^2 \mathbf{Q}_\perp   \left| F_{str}(\mathbf{p}_\perp, p_3, \mathbf{Q}_\perp, Q_z) \right|^2=E_p^*\frac{d^3\sigma^*}{d^3k^*},
\end{equation}
where  $\mathbf{p}_\perp$ and $p_3$ are the transverse and longitudinal momentum components respectively, and $\mathbf{Q}_\perp$ is the transferred transverse momentum, with $E_p^*$ being the proton energy expressed through their momenta in the deuteron rest frame.

Unlike in~\cite{Kobushkin2008}, we do not set to zero the transverse components of the proton-neutron pair $\mathbf{p}_\perp$ nor the longitudinal component of the momentum transfer $Q_z$, which enables investigation of the contribution from these kinematic factors.

As mentioned above, to obtain more general conclusions, the comparison with the experimental data  \cite{Ableev1983} \cite{Sitnik2019} was carried out for a broad class of deuteron potentials. Since for realistic deuteron potentials even when only the s-wave component is taken into account the invariant cross section acquires an extremely cumbersome analytical form (see Appendix D), we restrict ourselves here to a graphical representation of several characteristic consequences of including $Q_z$ and $\mathbf{p}_\perp$.

 The following results were obtained: for the single-Gaussian parametrization by Tartakovsky ~\cite{Tartakovsky2005} (Figs. 1-4); for the multi-Gaussian parametrization K2 ~\cite{Simenog} (Figs. 5-8); for the multi-Gaussian
 parametrization Av18  \cite{Wiringa1995} (Figs. 9 -14); and for the realistic Nijm-I potential  \cite{NijmI}  (Figs. 15, 16).

% Нами були отримані наступні результати: 

%\begin{itemize}
%\item Для одногаусової ''параметризації Тартаковського''  \cite{Tartakovsky2005}:

%\begin{itemize}
%\item For the single-Gaussian "Tartakovsky parametrization"~\cite{Tartakovsky2005}:

\begin{figure}[H]
\begin{center}
\includegraphics[scale=0.99]{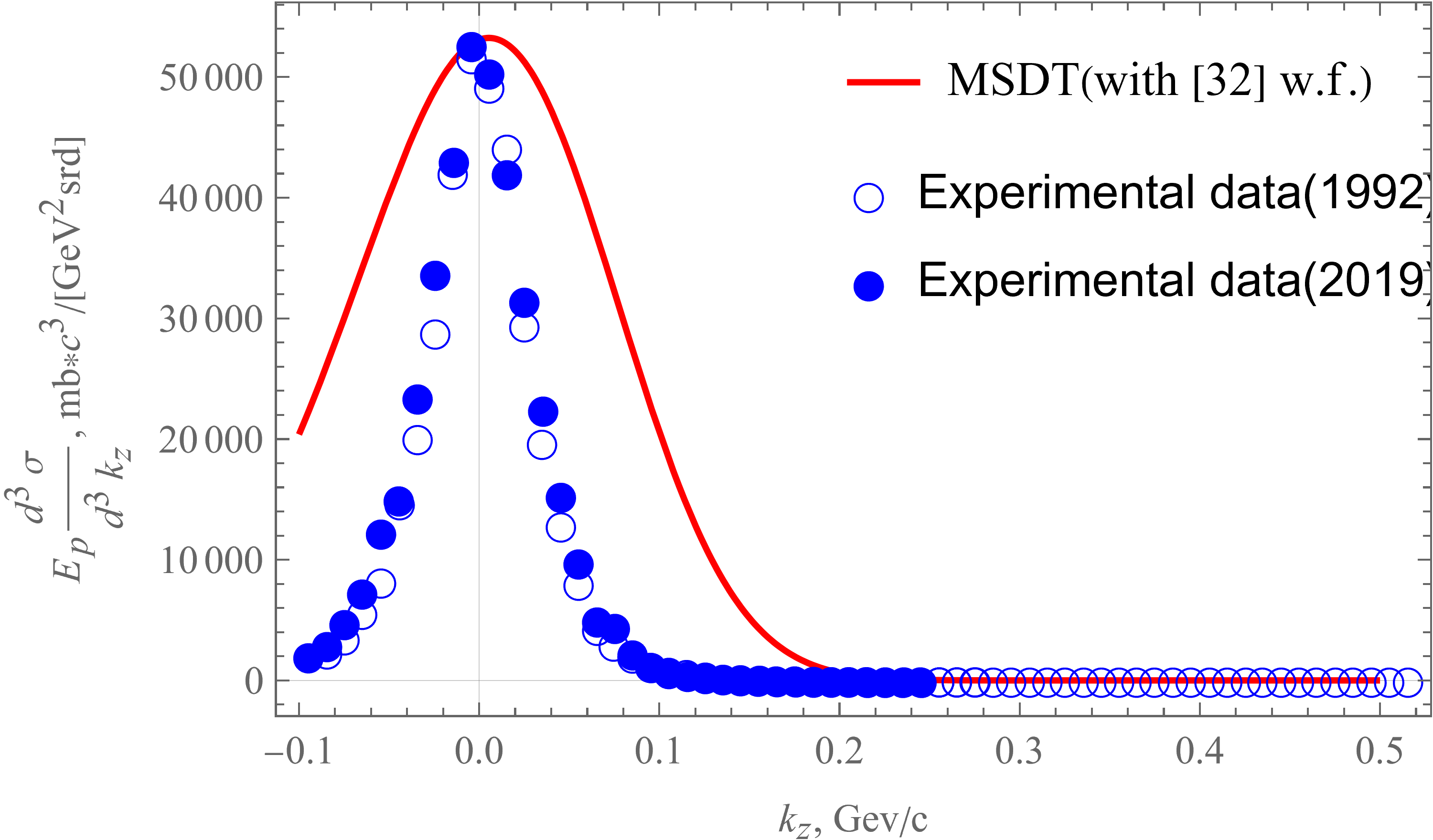}
\caption{
Dependence of \( E_p \, d^3 \sigma / d^3 k \) on \( k_z \), calculated using the single-Gaussian "Tartakovsky parametrization"~\cite{Tartakovsky2005}, for \( p_x = 0.00001 \, \text{GeV}/c \) and \( Q_z = 0.00001 \, \text{GeV}/c \). The point represents the experimental data approximation; the  continuous curve corresponds to the calculation based on the Glauber–Sitenko multiple scattering diffraction theory (MSDT).}
\label{Tart00001}
\end{center}
\end{figure}
%$\lambda -> \sqrt{.267}, \sigma_N -> 400., \rho_N -> 0.1, pd -> 9.1, \beta_N -> \sqrt{0.96}, N_s -> 1.0,  kx -> 0.01, Qz -> 0.007????????????????????$

%ass = {\[Lambda] -> Sqrt[.267], \[Sigma]N -> 92.5, \[Rho]N -> 0.001,   pd -> 9.1, \[Beta]N -> Sqrt[0.27], Ns -> 1.0, mp -> 938.272/1000,   Md -> 1875.6/1000, kx -> 0.00001, Qz -> 0.00001}
\begin{figure}[H]
\begin{center}
\includegraphics[scale=0.99]{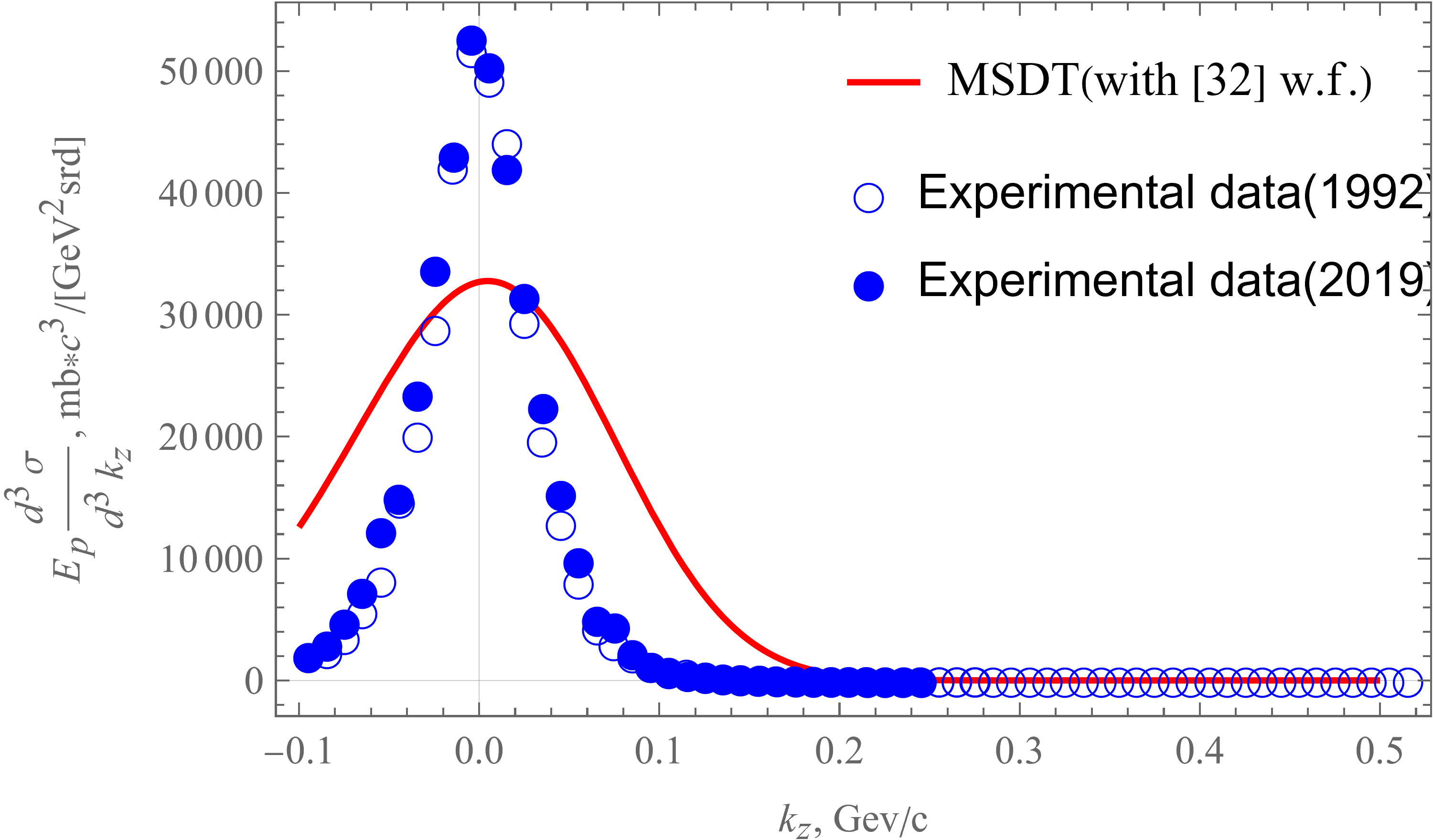}
\caption{
Dependence of \( E_p \, \frac{d^3 \sigma}{d^3 k} \) on \( k \), calculated using the single-Gaussian "Tartakovsky parametrization"~\cite{Tartakovsky2005}, for \( p_x = 0.5 \, \text{GeV}/c \) and \( Q_z = 0.00001 \, \text{GeV}/c \)}
%Залежність $E_p d^3 \sigma/d^3 k_z$ від $k_z$, отримана на х.ф. одногаусової ''параметризації Тартаковського''при $p_x=0.5$  GeV/c та $Q_z=0.00001$  GeV/c }
\label{Tart05}
\end{center}
\end{figure}

%ass = {\[Lambda] -> Sqrt[.267], \[Sigma]N -> 92.5, \[Rho]N -> 0.001,   pd -> 9.1, \[Beta]N -> Sqrt[0.27], Ns -> 1.0, mp -> 938.272/1000,   Md -> 1875.6/1000, kx -> 0.5, Qz -> 0.00001}

\begin{figure}[H]
\begin{center}
\includegraphics[scale=0.99]{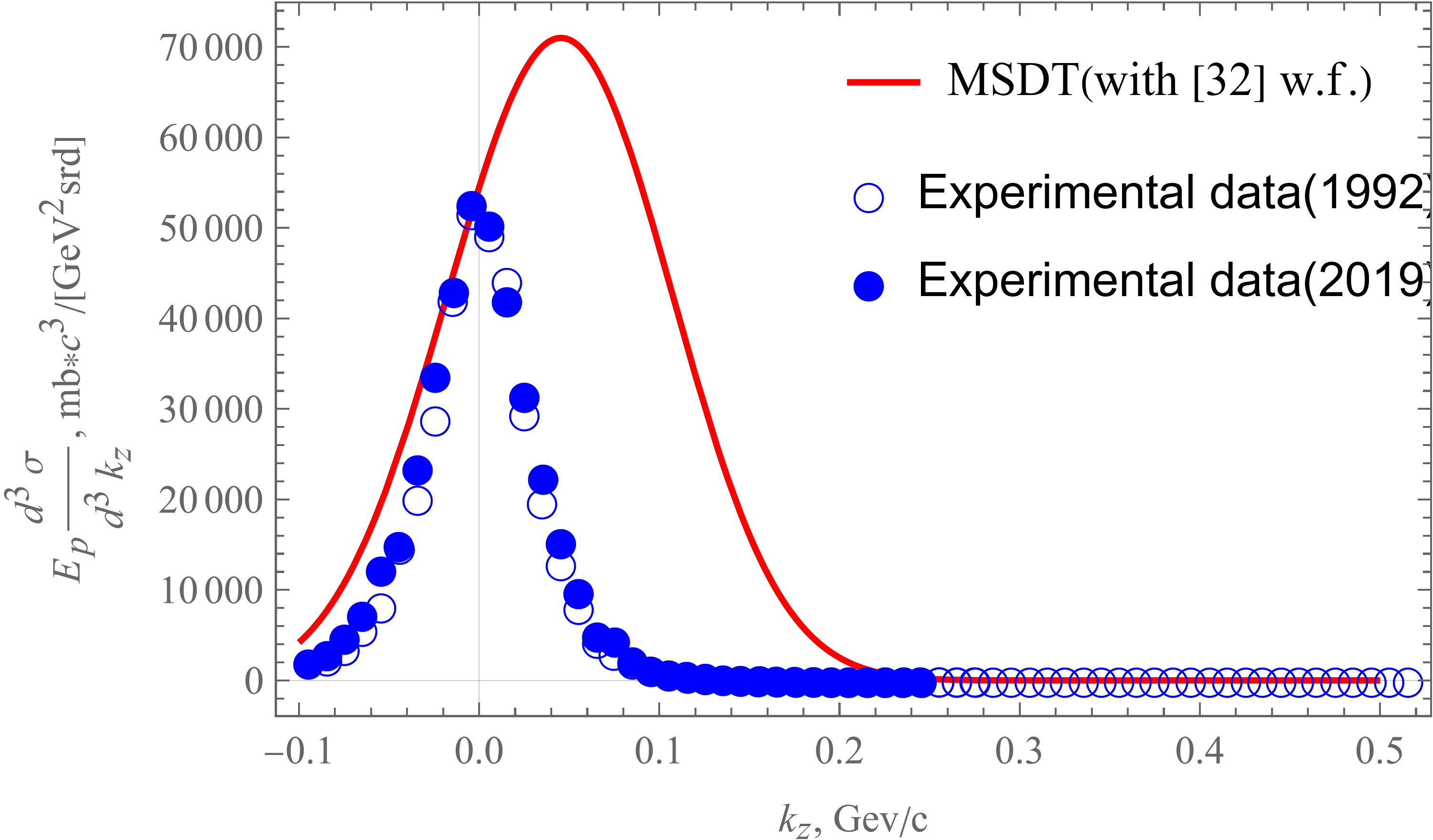}
\caption{
Dependence of \( E_p \, \frac{d^3 \sigma}{d^3 k} \) on \( k_z \), calculated using the single-Gaussian "Tartakovsky parametrization"~\cite{Tartakovsky2005}, for \( p_x = 0.00001 \, \text{GeV}/c \) and \( Q_z = 0.5 \, \text{GeV}/c \).}
%Залежність $E_p d^3 \sigma/d^3 k_z$ від $k_z$, отримана на х.ф. одногаусової ''параметризації Тартаковського''при $p_x=0.00001$ GeV/c та $Q_z=0.5$  GeV/c}
\label{TartQ03}
\end{center}
\end{figure}
$\lambda -> \sqrt{.267}, \sigma_N -> 40., \rho_N -> 0.1, pd -> 9.1, \beta_N -> \sqrt{0.96}, N_s -> 1.0, mp -> 938.272/1000, Md -> 1875.6/1000, kx -> 0.01, Qz -> 0.07$
%ass = {\[Lambda] -> Sqrt[.267], \[Sigma]N -> 40., \[Rho]N -> 0.01,  pd -> 9.1, \[Beta]N -> Sqrt[0.017], Ns -> 1.0, mp -> 938.272/1000,  Md -> 1875.6/1000, kx -> 0.0001, Qz -> 0.3}

\begin{figure}[H]
\begin{center}
\includegraphics[scale=0.99]{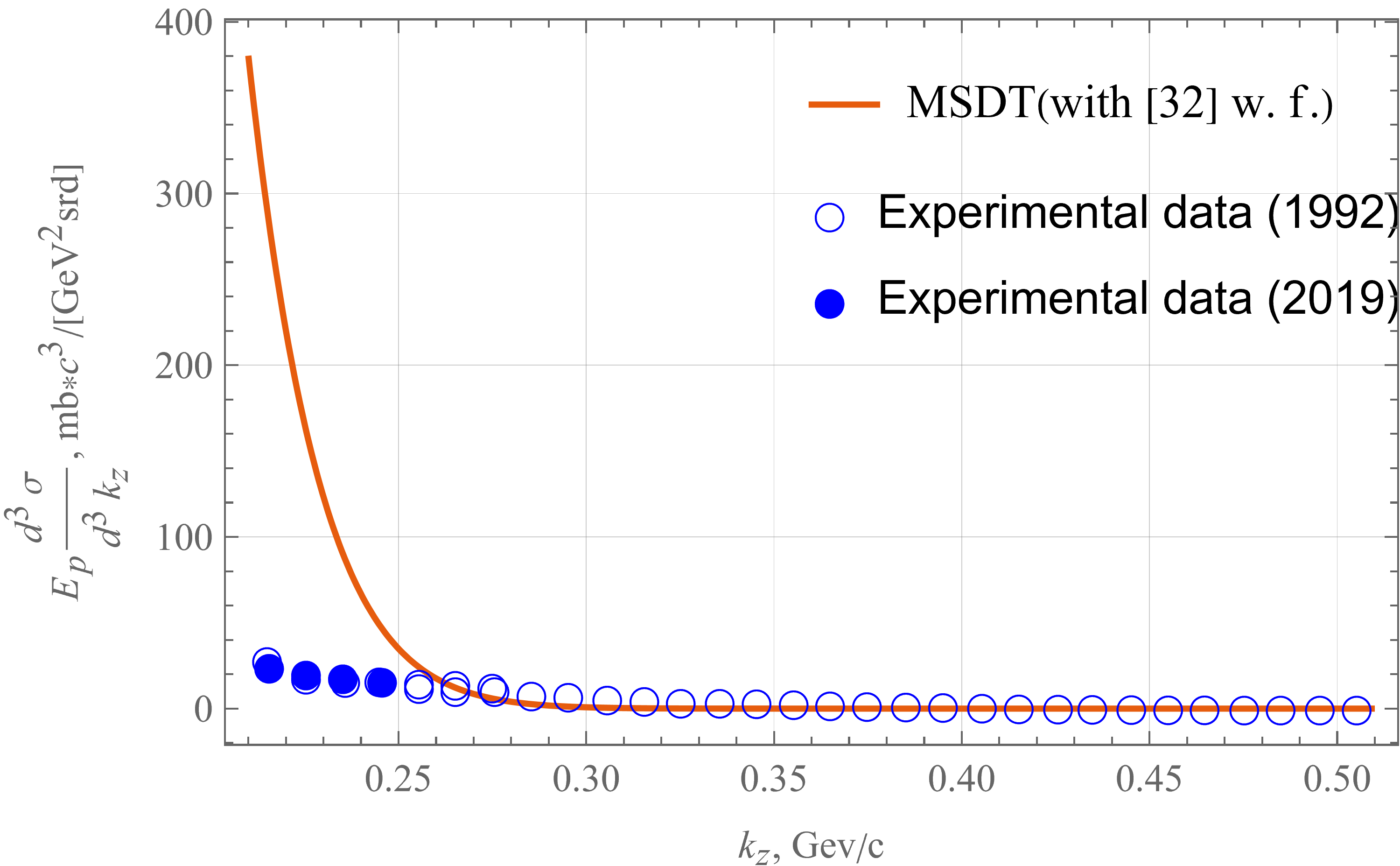}
\end{center}
\caption{ 
Within the framework of the Glauber–Sitenko model using wave functions based on the single-Gaussian parametrization~\cite{Tartakovsky2005}, the existence of dibaryon states with energies above 0.25~GeV/$c$ is not supported. The calculation was performed for \( p_x = 0.00001 \)~GeV/$c$, \( Q_z = 0.00001 \)~GeV/$c$}
% У рамках моделі Глаубера-Ситенка з хвильовими функціями, побудованими на основі одногаусової параметризації  \cite{Tartakovsky2005}, існування дібаріонних станів з енергіями більшими за  0.25GeV/c не знаходить підтвердження.   Розрахунок проведено при $p_x=0.00001$ GeV/c,   $Q_z=0.00001$ GeV/c }
\label{d(2880)TpQ-015}
\end{figure}

%\item For the multi-Gaussian S-wave parametrization K2~\cite{Simenog}:
%Для багатогаусової S-хвильової параметризації К2 \cite{Simenog}:
 
\begin{figure}[H]
\begin{center}
\includegraphics[scale=0.99]{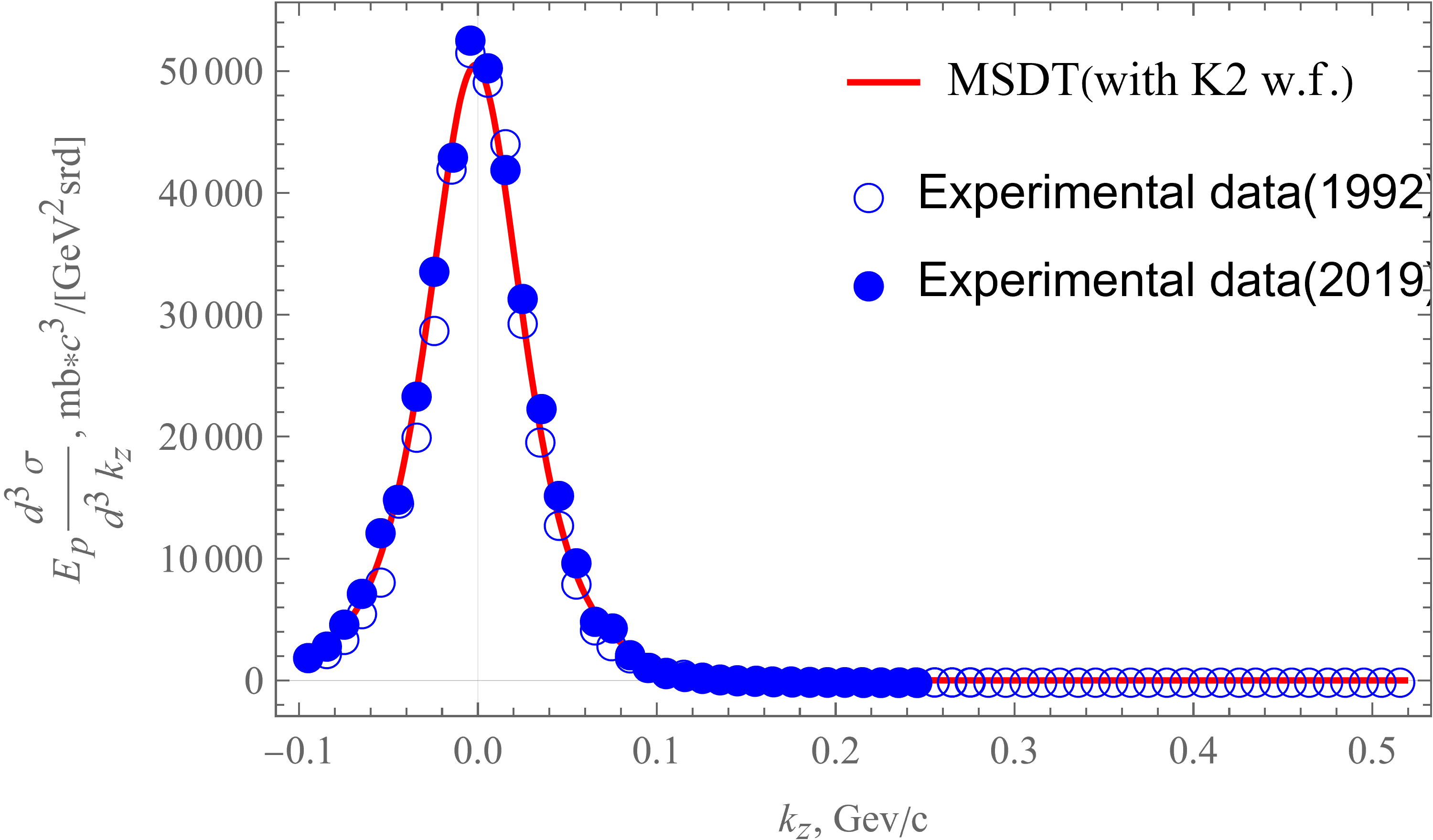}
\caption{
Dependence of \( E_p \, d^3 \sigma / d^3 k \) on \( k_z \), obtained using the wave function of the multi-Gaussian "K2 parametrization" at \( p_x = 0.000015 \)~GeV/\( c \), \( Q_z = -0.015 \)~GeV/\( c \)
}
%Залежність $E_p d^3 \sigma/d^3 k_z$ від $k_z$, отримана на х.ф. багатогаусової ''параметризації К2'' при $p_x=0.000015$ GeV/c $Q_z=-0.015$  GeV/c}
\label{K2p00015}
\end{center}
\end{figure}
%$\lambda -> \sqrt{.267}, \sigma_N -> 400., \rho_N -> 0.1, pd -> 9.1, \beta_N -> \sqrt{0.96}, N_s -> 1.0, mp -> 938.272/1000, Md -> 1875.6/1000, kx -> 0.01, Qz -> 0.007$
%s = {\[Sigma]N -> 194.5, \[Rho]N -> 0.001,   pd -> 9.1, \[Beta]N -> Sqrt[5.8500^2], Ns -> 1.0};
%d\[Sigma]d3kdc =  Table[{kz,    1/(2 \[Pi])^3 Sqrt[   mp^2 /. mp ->   938.272/1000 + 0.00015^2 + (scc[0.00015, kz] /. as2)^2]    Activate[   d\[Sigma]d3kd[0.00015, 0, lsc[0.00015, kz] /. as2, -0.015] /.    as // N]}, {kz, -0.10, 0.50, 0.01}]

%\begin{figure}
%\begin{center}
%\includegraphics[scale=0.95]{K2p_0003}
%\caption{
%Залежність $E_p d^3 \sigma/d^3 k_z$ від $k_z$, отримана на х.ф. багатогаусової ''параметризації К2'' при $k_x=0.003$ GeV/c  $Q_z=0.0001$  GeV/c}
%\label{K2p0003}
%\end{center}
%\end{figure}

% \begin{figure}
%\begin{center}
%\includegraphics[scale=0.95]{K2pQ_-03point}
%\caption{
%Залежність $E_p d^3 \sigma/d^3 k_z$ від $k_z$, отримана на х.ф. багатогаусової ''параметризації К2'' при $k_x=0.00015$ GeV/c  $Q_z=-0.03$  GeV/c}
%\label{K2pQ-03}
%\end{center}
%\end{figure}

\begin{figure}[H]
\begin{center}
\includegraphics[scale=0.99]{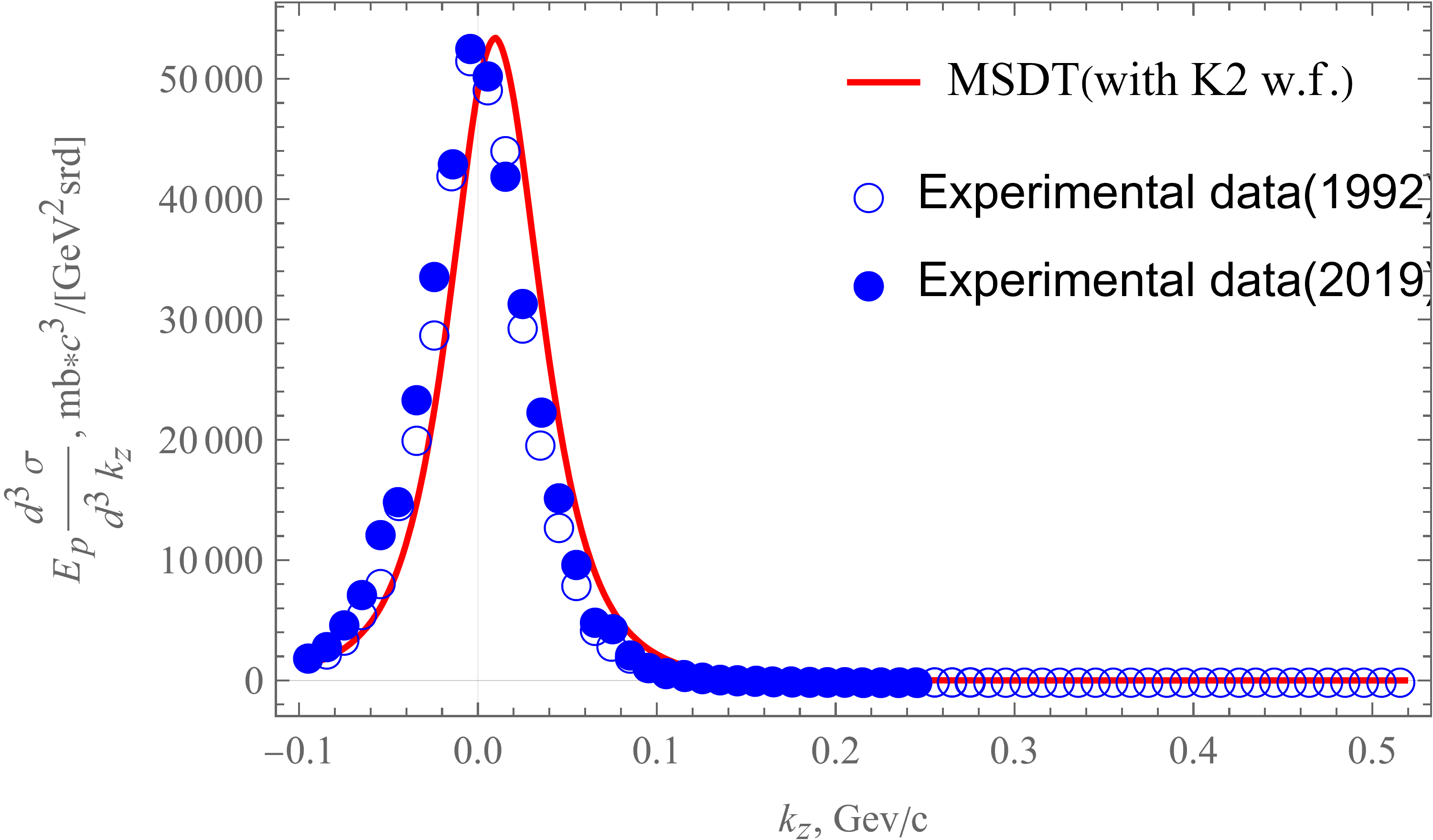}
\caption{
%Залежність $E_p d^3 \sigma/d^3 k_z$ від $k_z$, отримана на х.ф. багатогаусової ''параметризації К2'' при $p_x=0.000015$ GeV/c,   $Q_z=0.05$ GeV/c 
Dependence of \( E_p \, d^3 \sigma / d^3 k\) on \( k_z \), obtained using the wave function of the multi-Gaussian "K2 parametrization" at \( p_x = 0.000015 \)~GeV/\( c \), \( Q_z = 0.05 \)~GeV/\( c \)
}
\label{K2pQ05}
\end{center}
\end{figure}

\begin{figure}[H]
\begin{center}
\includegraphics[scale=0.99]{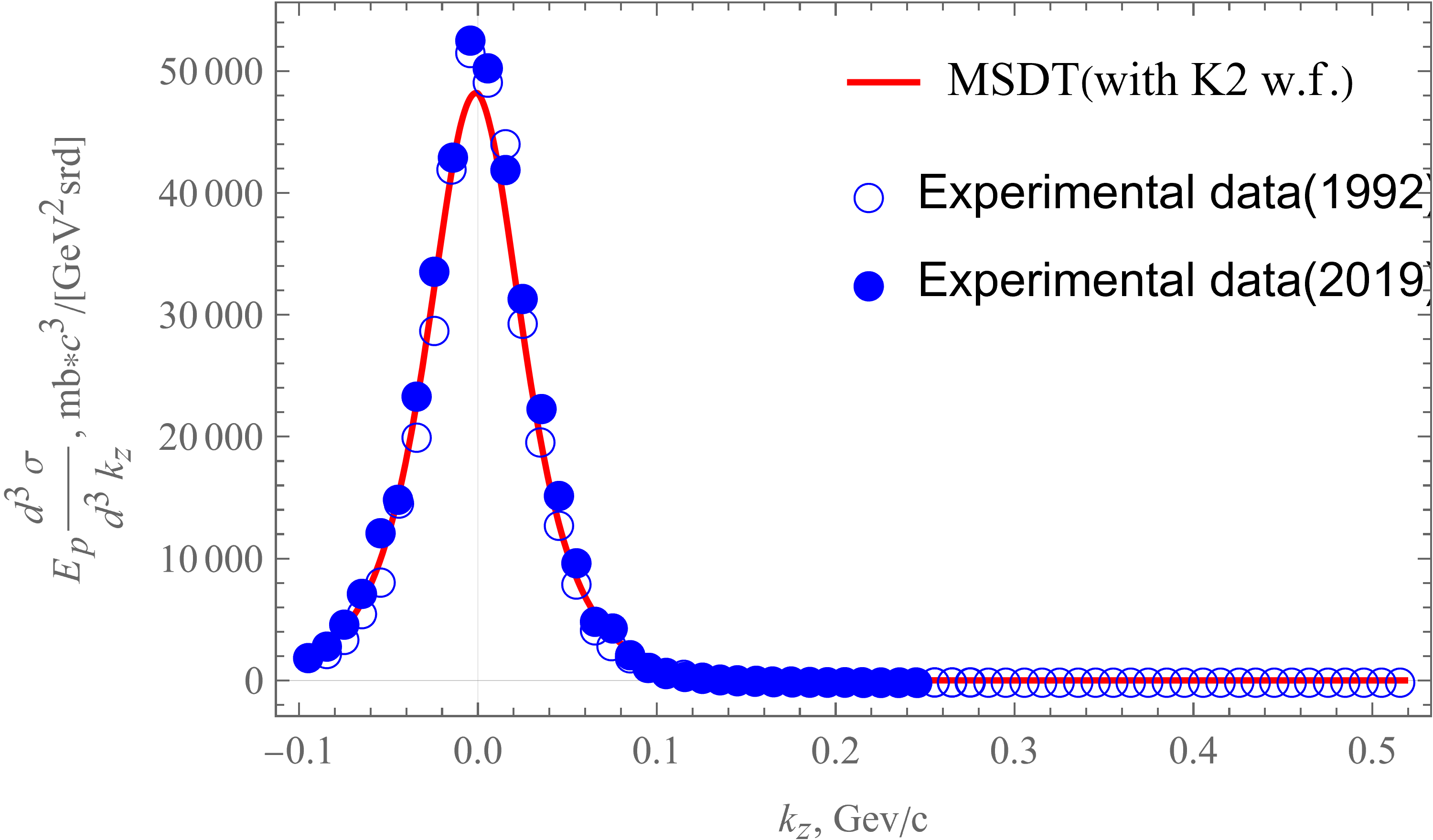}
\caption{
Dependence of \( E_p \, d^3 \sigma / d^3 k \) on \( k_z \), obtained using the wave function of the multi-Gaussian "K2 parametrization" at \( p_x = 0.05 \)~GeV/\( c \), \( Q_z =- 0.015 \)~GeV/\( c \)}
%Залежність $E_p d^3 \sigma/d^3 k_z$ від $k_z$, отримана на х.ф. багатогаусової ''параметризації К2'' при $p_x=0.05$ GeV/c  $Q_z=-0.015$  GeV/c}
\label{K2p005}
\end{center}
\end{figure}

\begin{figure}[H]
\begin{center}
\includegraphics[scale=0.99]{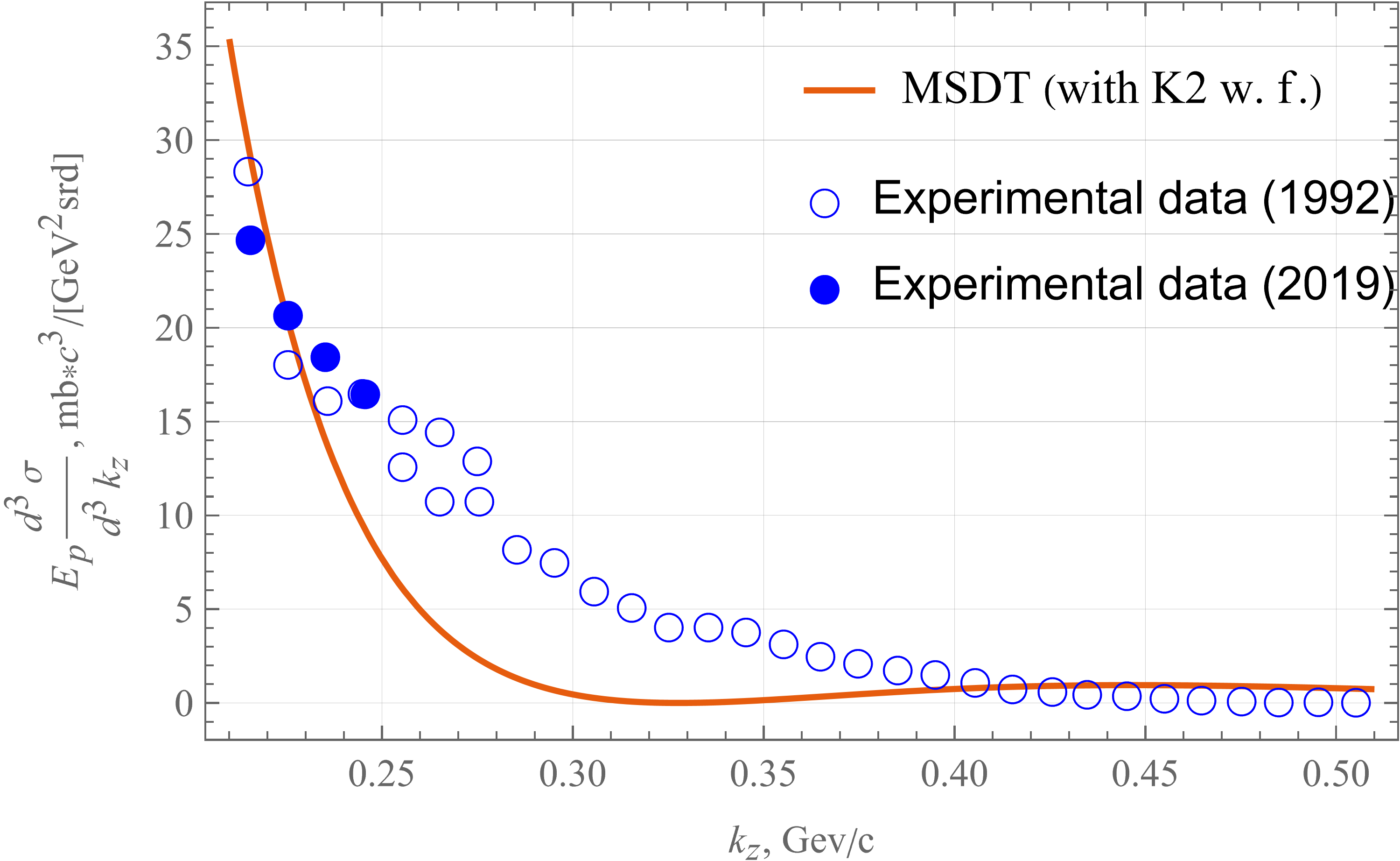}
\end{center}
\caption{ 
Within the framework of the Glauber–Sitenko model using wave functions based on the K2 potential, the existence of the dibaryon state $d^*(2380)$ is not supported. However, the model does not exclude the possibility of other dibaryon resonances forming at lower momenta (0.23–0.40~GeV/\textit{c}), as well as the contribution of states involving S-quarks in this lower-energy region. The calculation was performed for \( p_x = 0.000015 \)~GeV/\textit{c}, \( Q_z = -0.015 \)~GeV/\textit{c}}
% У рамках моделі Глаубера-Ситенка з хвильовими функціями, побудованими на основі потенціалу K2, існування дібаріонного стану $d^*(2380)$ не знаходить підтвердження. Однак модель не виключає можливості утворення інших дібаріонних резонансів при нижчих енергіях (0.23-0.40 GeV/c ), а також внеску %станів із участю S-кварків у цій, більш низькоенергетичній області.   Розрахунок проведено при $p_x=0.000015$ GeV/c,   $Q_z=-0.015$ GeV/c }
\label{d(2880)K2pQ-015}
\end{figure}

% \item For the Argonne-type potential AV18 \cite{Wiringa1995}, taking into account only the S-wave:

\begin{figure}[H]
\begin{center}
\includegraphics[scale=0.99]{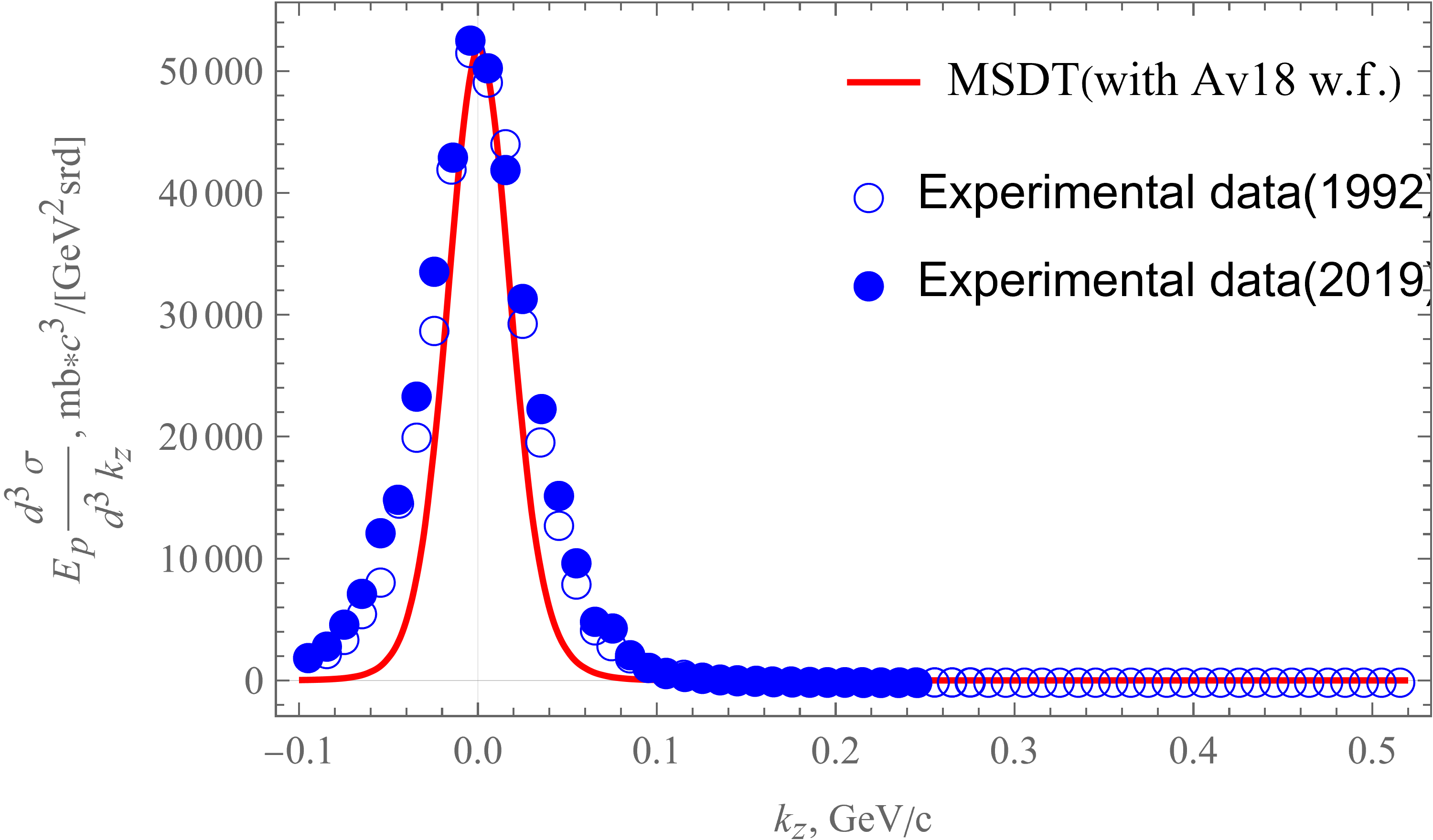}
\caption{
Dependence of $E_p d^3 \sigma/d^3 k$ on $k_z$, obtained using the multi-Gaussian wave function of the ''Av18 parametrization' at $p_x = 0.00001$ GeV/c, $Q_z = -0.5$ GeV/c}
\label{Av180001}
\end{center}
\end{figure}
%as = {\[Sigma]N -> 38.0, \[Rho]N -> 0.00001,    pd -> 9.1, \[Beta]N -> Sqrt[1950.5^2], Ns -> 0.999};

\begin{figure}[H]
\begin{center}
\includegraphics[scale=0.99]{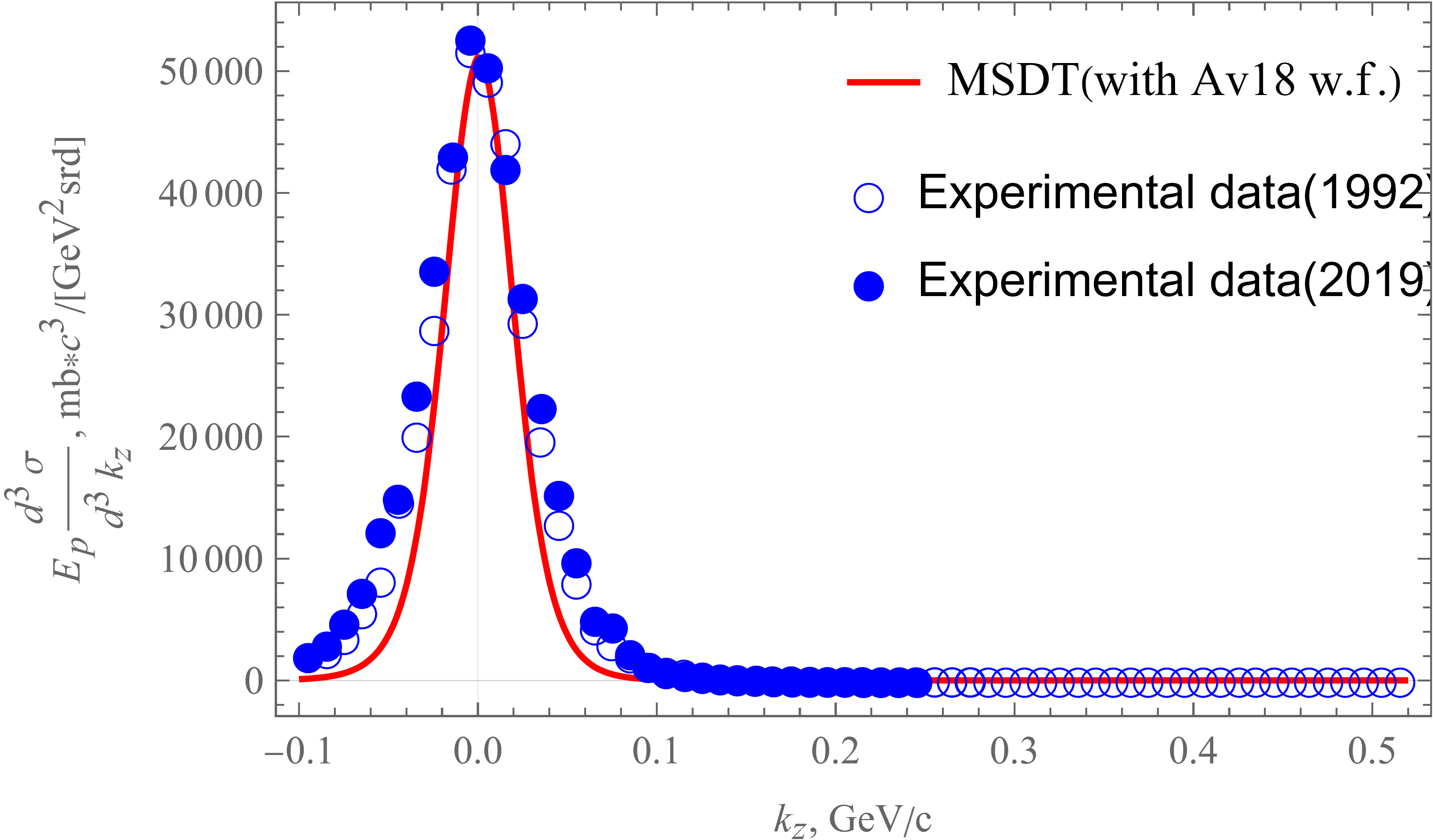}
\caption{
Dependence of $E_p d^3 \sigma/d^3 k$ on $k_z$, obtained using the multi-Gaussian wave function of the ''Av18 parametrization' at $p_x = 0.00001$ GeV/c, $Q_z = 0.00001$ GeV/c}
%Залежність $E_p d^3 \sigma/d^3 k_z$ від $k_z$, отримана на х.ф. багатогаусової ''параметризації Av18'' при $p_x=0.00001$ GeV/c  та  $Q_z=0.00001$ GeV/c}
\label{Av18012}
\end{center}
\end{figure}

\begin{figure}[H]
\begin{center}
\includegraphics[scale=0.99]{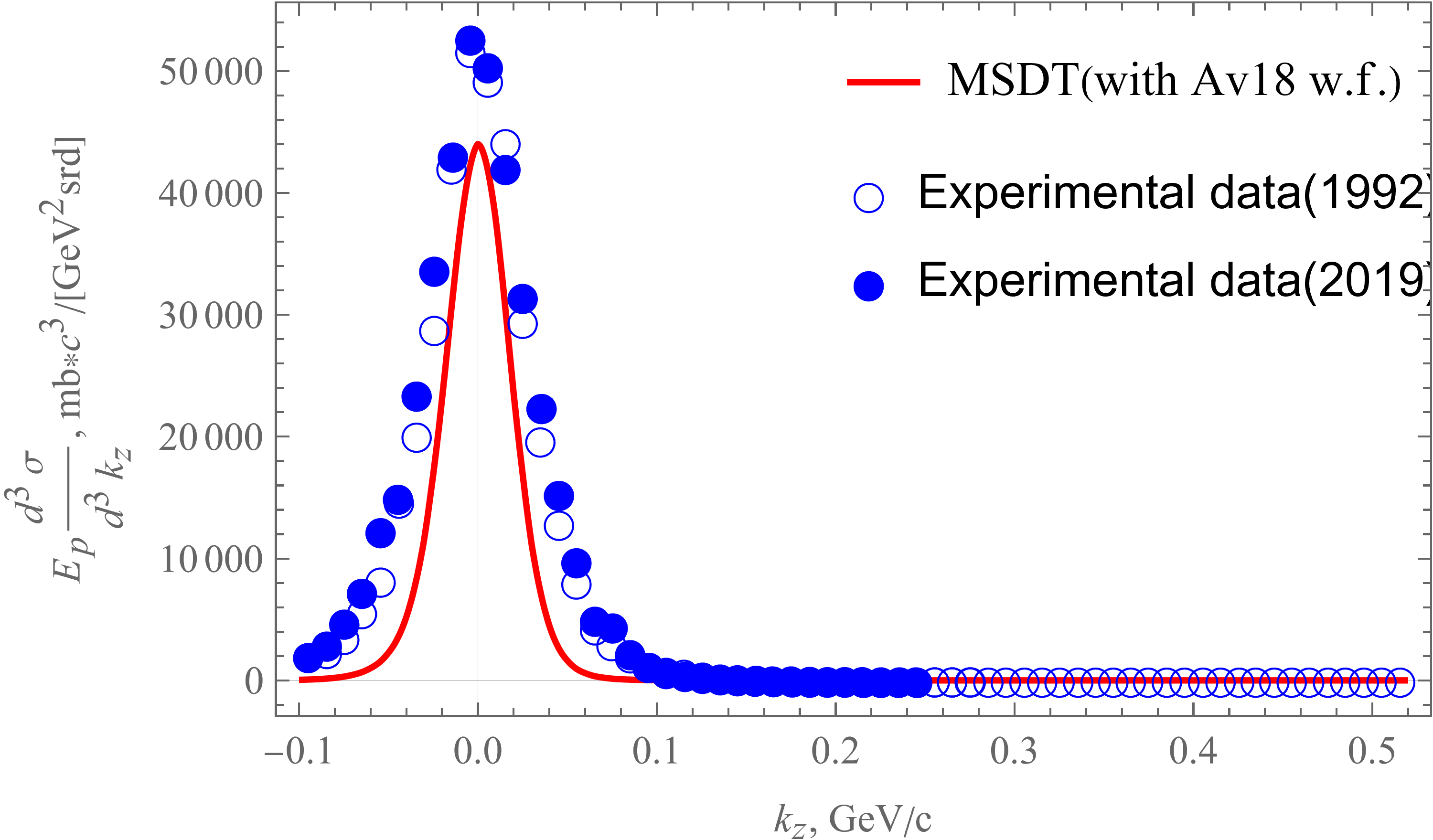}
\caption{
Dependence of $E_p d^3 \sigma/d^3 k$ on $k_z$, obtained using the multi-Gaussian wave function of the ''Av18 parametrization' at $p_x = 0.05$ GeV/c, $Q_z = 0.5$ GeV/c}
%Залежність $E_p d^3 \sigma/d^3 k_z$ від $k_z$, отримана на х.ф. багатогаусової ''параметризації Av18'' при $p_x=0.05$ GeV/c та $Q_z=0.5$ GeV/c}
\label{Av18005}
\end{center}
\end{figure}

\begin{figure}[H]
\begin{center}
\includegraphics[scale=0.99]{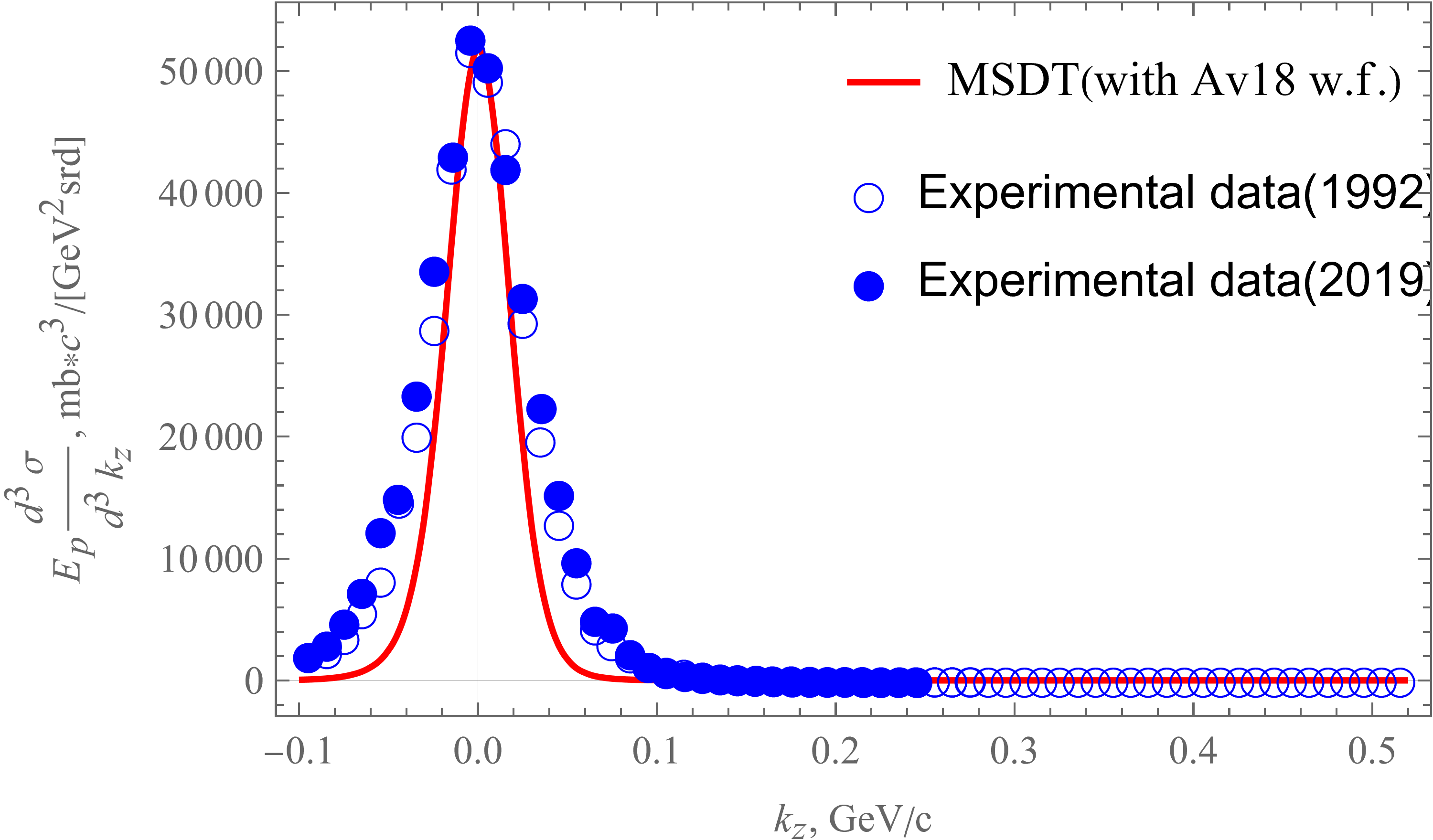}
\caption{
Dependence of $E_p d^3 \sigma/d^3 k$ on $k_z$, obtained using the multi-Gaussian wave function of the ''Av18 parametrization' at $p_x = 0.00001$ GeV/c, $Q_z = 0.5$ GeV/c}
%Залежність $E_p d^3 \sigma/d^3 k_z$ від $k_z$, отримана на х.ф. багатогаусової ''параметризації Av18'' при $p_x=0.00001$ GeV/c  та $Q_z=0.5$ GeV/c}
\label{Av18Q05}
\end{center}
\end{figure}

\begin{figure}[H]
\begin{center}
\includegraphics[scale=0.99]{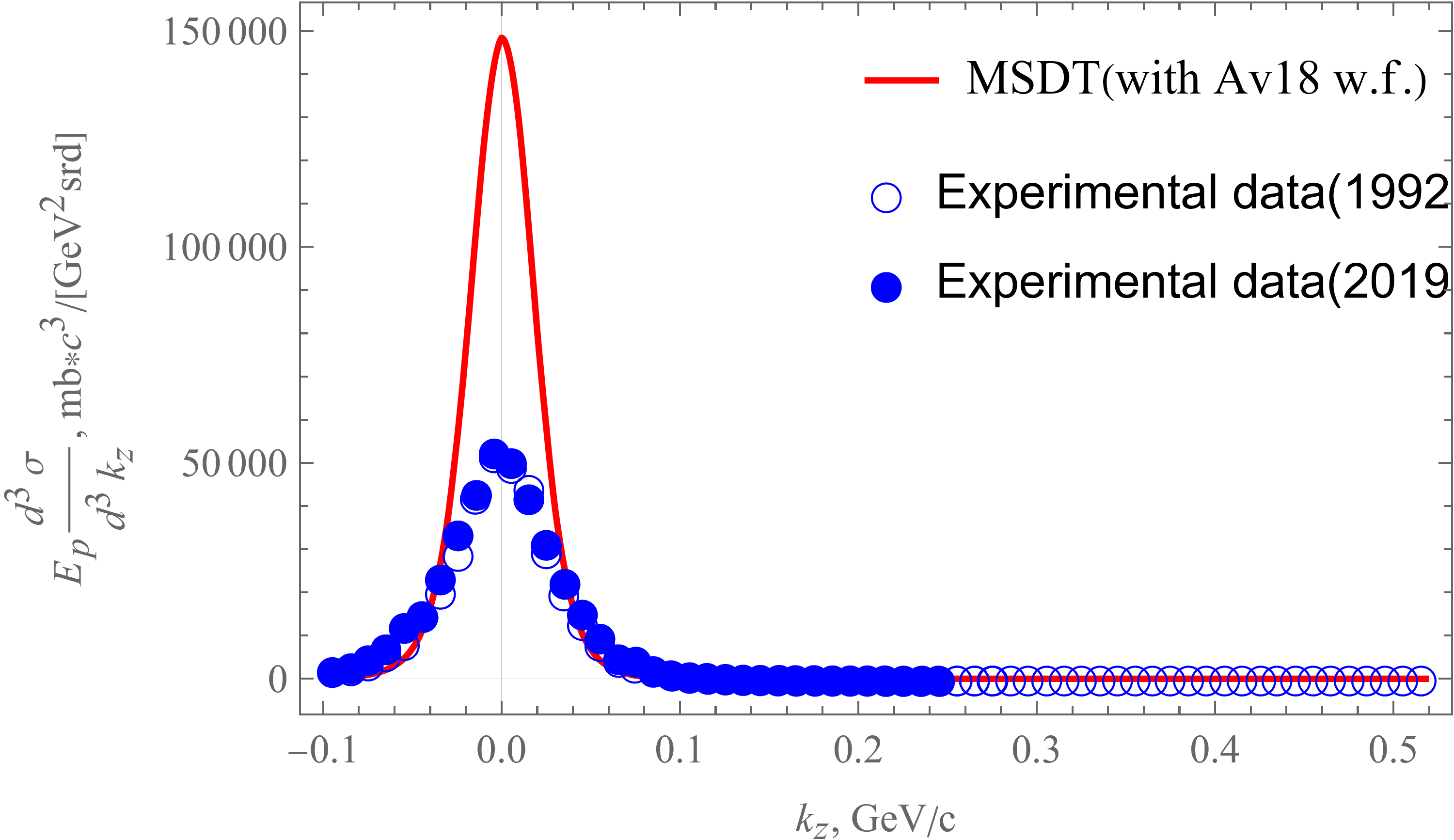}
\caption{
Dependence of $E_p d^3 \sigma/d^3 k$ on $k_z$, obtained using the multi-Gaussian wave function of the ''Av18 parametrization' at $p_x = 0.12$ GeV/c, $Q_z = 0.00001$ GeV/c}
\label{Av1800001}
\end{center}
\end{figure}
%as = {\[Sigma]N -> 38.0, \[Rho]N -> 0.00001,    pd -> 9.1, \[Beta]N -> Sqrt[1950.5^2], Ns -> 0.999};

\begin{figure}[H]
\begin{center}
\includegraphics[scale=0.99]{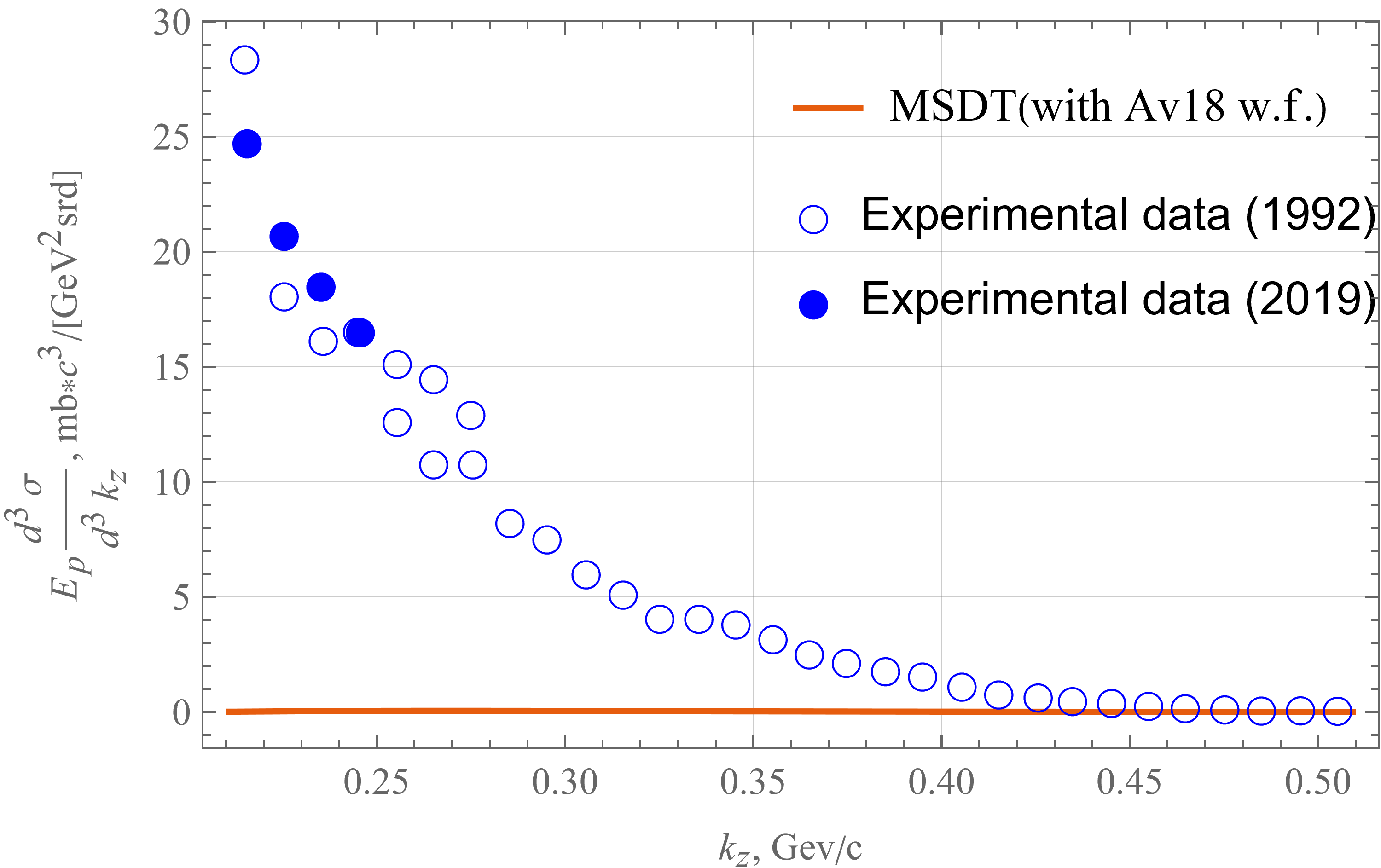}
\end{center}
\caption{
Within the framework of the Glauber-Sitenko model using wave functions constructed from the Av18 potential, the existence of the dibaryon state $d^*(2380)$ is not supported. However, the model does not exclude the possibility of forming other dibaryon resonances at energies below 0.40 GeV/c, as well as contributions from states involving $S$-quarks in this lower-energy region. The calculation was performed for $p_x = 0.00001$–$0.12$ GeV/c and $Q_z = 0.00001$–$0.5$ GeV/c}
\label{d(2880)Av18-5}
\end{figure}
%as = {\[Sigma]N -> 38.0, \[Rho]N -> 0.00001,    pd -> 9.1, \[Beta]N -> Sqrt[1950.5^2], Ns -> 0.999};
%\item For the non-local Nijm-I potential \cite{NijmI} including both $S$- and $D$-waves (see Fig.~\ref{Nijm05}):

\begin{figure}[H]
\begin{center}
\includegraphics[scale=0.99]{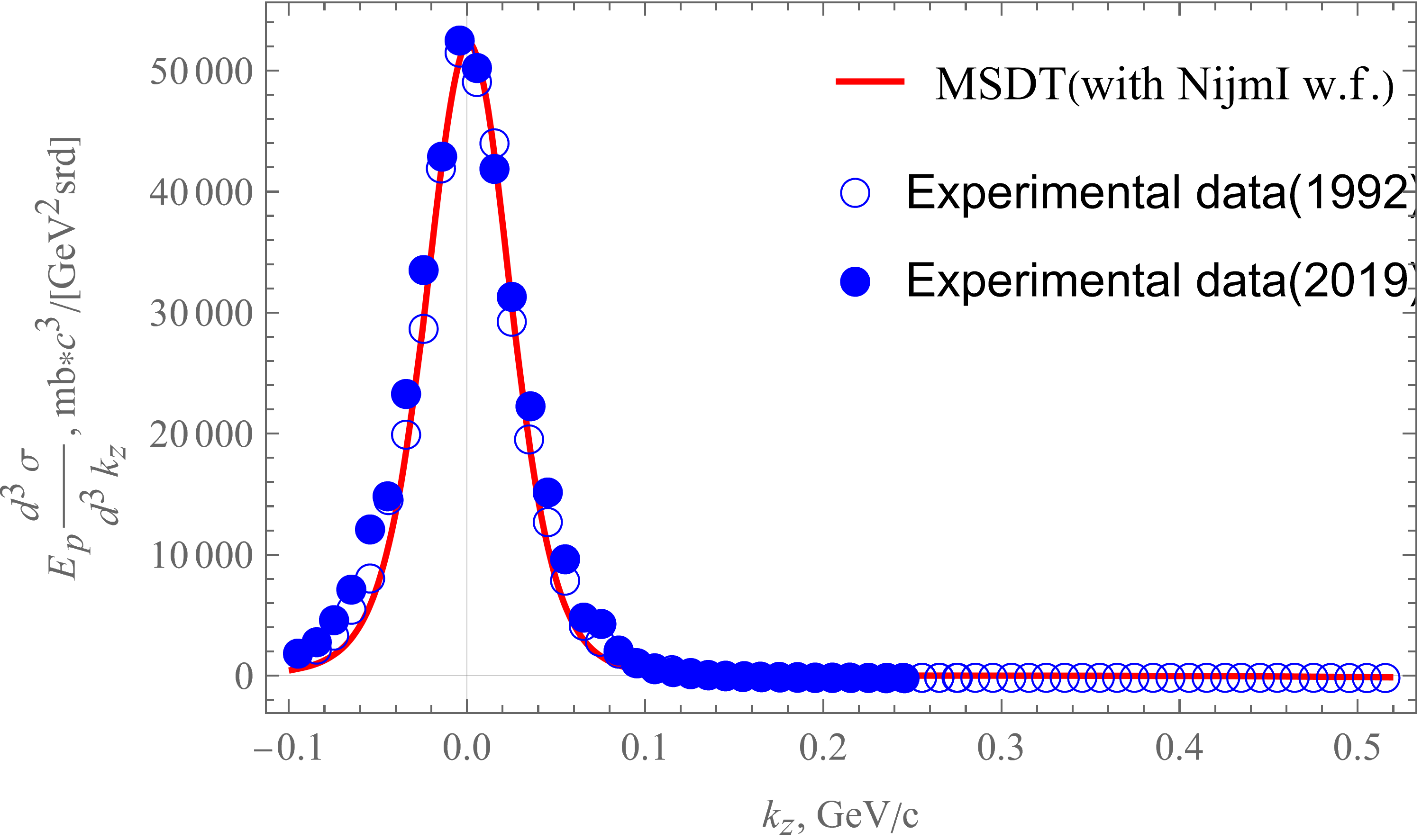}
\caption{
Dependence of $E_p d^3 \sigma/d^3 k$ on $k_z$, obtained using the multi-Gaussian wave function of the ''Nijm-I parametrization'' at $p_x = 0.145$ GeV/c, $Q_z = -0.01$ GeV/c}
%Залежність $E_p d^3 \sigma/d^3 k_z$ від $k_z$, отримана на х.ф. Nijm-I -потенціалу  при $p_x=0.145$ GeV/c та  $Q_z=-0.015$ GeV/c}
\label{Nijm05}
\end{center}
\end{figure}
%as = {\[Sigma]N -> 50., \[Rho]N -> 0.001,   pd -> 9.1, \[Beta]N -> Sqrt[480.^2], \[Alpha] -> 1};
%\end{itemize}

\begin{figure}[H]
\begin{center}
\includegraphics[scale=0.99]{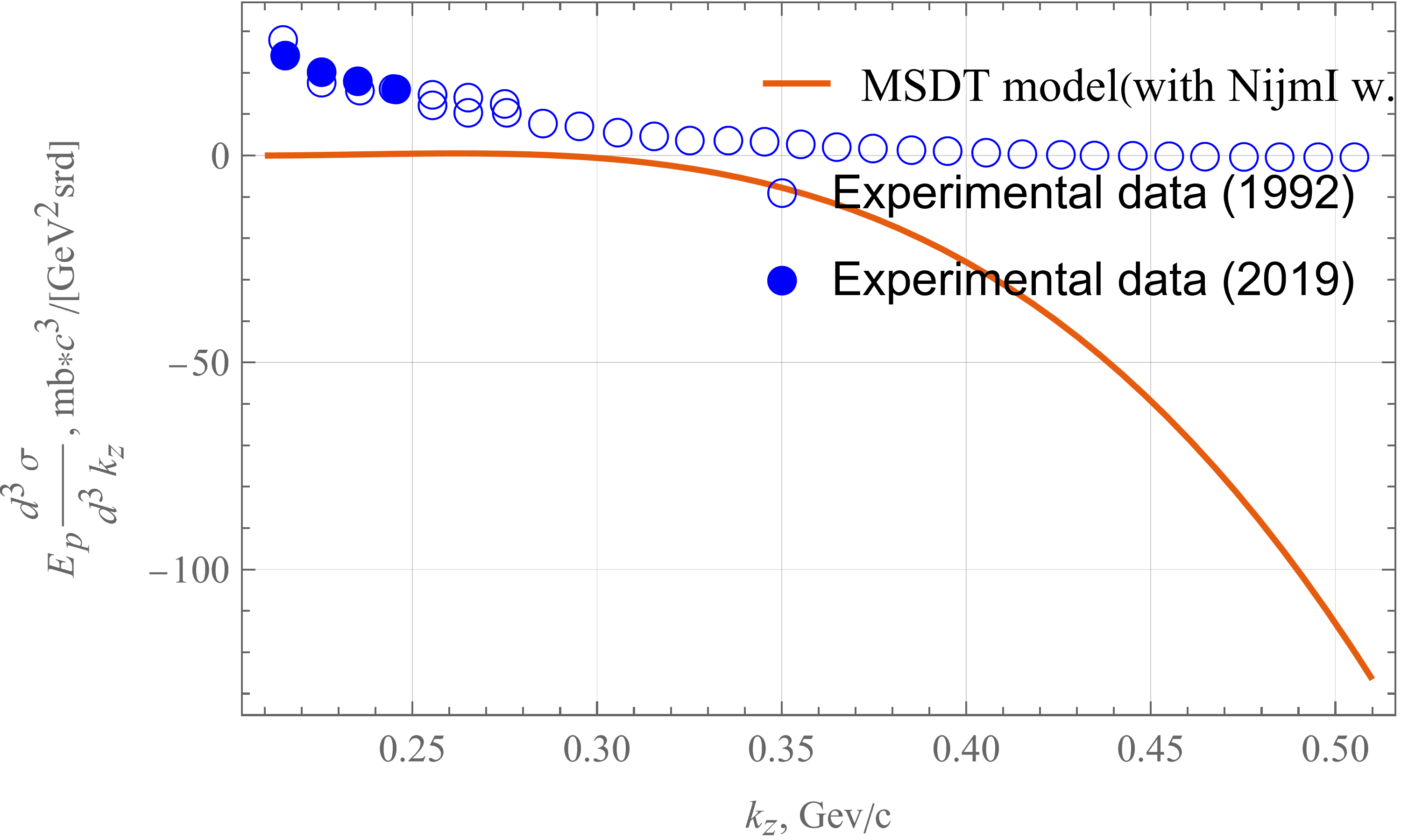}
\end{center}
\caption{
Within the Glauber–Sitenko model using wave functions based on the NijmI potential, the existence of the dibaryon state $d^*(2380)$ is not excluded. The model also allows for the possible formation of other dibaryon resonances at momenta below 0.5~GeV/c, as well as the contribution of states involving strange (S) quarks in this low-energy region. The calculation was performed for $p_x = 0.145$~GeV/c and $Q_z = -0.015$~GeV/c
}
\label{Nijmd(2880)}
\end{figure}

\section{Conclusion}
\begin{itemize}
     \item A comparison of Figures~\ref{Tart00001}, \ref{Tart05}, and \ref{TartQ03} leads to the conclusion that the experimental data cannot be adequately described using a single-Gaussian model, despite the fact that the overall functional dependence on the vectors $k_x$ and $Q_z$ remains unchanged. This indicates the necessity of refining the model, for instance, by introducing additional Gaussian components or by employing an alternative parameterization of the wave functions.

%     \item З порівнянь на малюнках  \ref{Tart00001}, \ref{Tart05}, \ref{TartQ03} можна зробити висновок, що  експериментальні дані не можуть бути адекватно описані за допомогою одногаусової моделі, незважаючи на те, що загальний функціональний вигляд залежності від векторів $k_x$ та   $Q_z$ %залишається незмінним. Це вказує на необхідність уточнення моделі, наприклад, через введення додаткових гаусових компонент або використання іншої параметризації хвильових функцій.
     \item While comparison with experiment suggests that small longitudinal momenta may be considered to explain the shift in the proton yield maximum (observed at proton momenta close to half the deuteron momentum in the laboratory frame), it does not indicate any significant influence of the longitudinal momentum on the dibaryon "enhancement" region in the differential cross section.
%    \item Хоча порівняння з експериментом свідчить про можливість врахування невеликих повздовжніх імпульсів для пояснення зсуву області максимуму виходу протона (що спостерігається при імпульсах протона, близьких до половини імпульсу дейтрона в лабораторній системі), однак воно не вказує на суттєвий %вплив повздовжнього імпульсу на область дібаріонного «напливу» у диференційному перерізі.

   \item  Within the framework of the Glauber–Sitenko model using wave functions based on the K2 potential, the existence of the dibaryon state $d^*(2380)$ is not supported. However, the model does not exclude the possibility of other dibaryon resonances forming at lower momenta (0.23–0.40~GeV/\textit{c}), as well as the contribution of states involving S-quarks in this lower-energy region. The calculation was performed for \( p_x = 0.000015 \)~GeV/\textit{c}, \( Q_z = -0.015 \)~GeV/\textit{c}
   
    \item  Within the framework of the Glauber-Sitenko model using wave functions constructed from the Av18 potential, the existence of the dibaryon state $d^*(2380)$ is not supported. However, the model does not exclude the possibility of forming other dibaryon resonances at energies below 0.40 GeV/c, as well as contributions from states involving $S$-quarks in this lower-energy region. The calculation was performed for $p_x = 0.00001$–$0.12$ GeV/c and $Q_z = 0.00001$–$0.5$ GeV/c
    
    \item Within the Glauber–Sitenko model using wave functions based on the NijmI potential, the existence of the dibaryon state $d^*(2380)$ is not excluded. The model also allows for the possible formation of other dibaryon resonances at momenta below 0.5~GeV/c, as well as the contribution of states involving strange (S) quarks in this low-energy region. The calculation was performed for $p_x = 0.145$~GeV/c and $Q_z = -0.015$~GeV/c
    
      \item The results obtained using wave functions from the K2, Av18, and Nijm-I potentials indicate that the observed enhancement ("bump") in the 300-500 MeV/c region of the invariant cross section in the antilaboratory frame, reported in \cite{Ableev1983}, cannot be explained within the framework of the Glauber–Sitenko multiple scattering theory (MSDT) - although partial overlap exists in K2 and Av18 starting from 400 MeV/c. This provides additional arguments that this "bump" is not related to the contribution from rescattered pions (whose effects should be effectively accounted for in realistic potentials and the profile function), but rather to quark effects. These findings support the conclusions presented in \cite{Ableev1983} and \cite{Kobushkin1982}, which suggest that the dominant mechanism in this energy range is the contribution from the S-quark component of the wave function or, according to an alternative interpretation, the contribution from a dibaryon state \cite{Sitenko2000}. 
      %This contradicts several later studies \cite{Braun1984, Braun1986} where similar effects were attributed exclusively to final-state pion exchange effects in the production of cumulative protons.
    
%    Отримані на основі  хвильових функцій потенціалів K2, Av18 та Nijm-I результати свідчать, що знайдений у роботі \cite{Ableev1983} підйом (''наплив''' ) у області 300-500~МеВ/с інваріантного перерізу в антилабораторній системі не пояснюється ДТБР(хоча у K2 та Av18 є часткове прекриття, починаючи з 400 %MeV/c), що дає додаткові аргументи тому, що цей ''наплив'' пов’язаний не з внеском перерозсіяних піонів, внесок яких мусить ефективно враховуватись у реалістичних потенціалах та профільній функції,  а з кварковими ефектами. Це підтверджує висновки, наведені в \cite{Ableev1983}, \cite{Kobushkin1982}, де %зазначено, що основним механізмом у цьому енергетичному діапазоні є внесок S-кваркової компоненти хвильової функції, або, згідно іншій інтерпретації, внеском дібаріонного стану \cite{Sitenko2000}. Це суперечить низці пізніших робіт \cite{Braun1984}, \cite{Braun1986}, де подібні ефекти пояснювались %виключно ефектами обміну піонами у кінцевому стані в народження кумулятивних протонів.
     
     \item Within the Glauber--Sitenko approximation, a comparison with experimental data on the differential cross section of deuteron breakup on a proton in the reaction $H(d,p)X$ was carried out, taking into account small longitudinal components of the transferred momentum and transverse components of the proton--neutron pair in the anti-laboratory frame. Preliminary estimates indicated a decrease in the cross section with increasing transverse momentum $\mathbf{p}_\perp$ and an increase in the cross section with increasing longitudinal component $Q_z$.
     
%     \item В наближенні Глаубера--Ситенка було проведено порівняння з експериментальними даними щодо диференційного перерізу розпаду дейтрона на протоні в реакції $H(d,p)X$ з урахуванням невеликих повздовжніх компонент переданого імпульсу та поперечних компонент пари протон--нейтрон в %антилабораторній системі координат. Попередні оцінки показали зменшення перерізу зі збільшенням поперечного імпульсу $\mathbf{p}_\perp $ та збільшення перерізу при збільшенні повздовжньої компоненти $Q_z$.
    \item To conclusively determine the influence of the longitudinal component of the transferred momentum $Q_z$ and transverse momenta in the antilaboratory frame $\mathbf{p}_\perp$, it is necessary to compute the corresponding Coulomb amplitudes, as they also depend on these momenta
%    Хоча, для $H(d,p)X$ завдяки відносно малому внеску кулонівська взаємодія  не так важлива. 
%    Для остаточного з'ясування впливу повздовжньої компоненти переданого імпульсу $Q_z$ та поперечних імпульсів в антилабораторній системі необхідно обчислити відповідні кулонівські амплітуди, оскільки вони також залежать від цих імпульсів. 
\end{itemize}

\section{Appendix A}
\addcontentsline{toc}{section}{Appendix A}

%\subsection*{Передумови}

Since in the anti-laboratory frame the following relation holds:
\[
k_3 = p_3^*,
\]
where \( p_3^* \) is the longitudinal momentum of the proton in the boosted (anti-laboratory) frame, in which
\[
p_3^* + n_3^* = 0 \quad \text{and} \quad p_d^* = 0,
\]
the Lorentz boost for \( p_3^* \) in this system takes the form:
\[
p_3^* = \gamma (p_3 - \beta E_p),
\]
where:
\[
\beta = \frac{p_d}{E_d}, \quad \gamma = \frac{E_d}{M_d}, \quad E_d = \sqrt{M_d^2 + p_d^2},
\]
\[
p_3 = \frac{1}{2} p_d + e_p, \quad E_p = \sqrt{M^2 + p_\perp^2 + p_3^2}.
\]

Here, \( p_3 \) is the longitudinal momentum of the proton in the laboratory frame, and \( e_p \) is the excess momentum that reflects the asymmetry in the momentum distribution \( p_d \) between the proton and the neutron. In the laboratory frame, the deuteron of mass \( M_d \) has momentum \( p_d \) and energy \( E_d \).
To find the boost velocity $\beta$ between frames, we used relativistic momentum transformation. In the boosted frame:

\begin{align}
p_d^* &= \gamma (p_d - \beta E_d) = 0 \nonumber \\
\Rightarrow \quad \beta &= \frac{p_d}{E_d} = \frac{p_d}{\sqrt{p_d^2 + M_d^2}} \nonumber \\
\gamma &= \frac{E_d}{M_d} = \frac{1}{\sqrt{1 - \beta^2}}=\frac{p_d}{E_d}
\end{align}

%\subsection*{Вираз для \( p_3^* \):}

Substituting \( p_3 \) and \( E_p \) into the formula for \( p_3^* \), we get:

\[
p_3^* = \gamma \left( \frac{1}{2} p_d + e_p - \beta \sqrt{M^2 + p_\perp^2 + \left( \frac{1}{2} p_d + e_p \right)^2} \right).
\]

Accordingly, the inverse Lorentz transformation takes the form:
\[
p_3 = \gamma( p_3^* \pm \beta \sqrt{(p_3^*)^2 + M^2 + p_\perp^2})= \gamma( p_3^* \pm \beta E_p^* )
\]
where \( E_p^* \) is the energy of the proton in the boosted frame.
The physical interpretation requires that when \( p_3^* = 0 \), we have
\[
p_3 = \beta \gamma \sqrt{M^2 + p_\perp^2} \sim \frac{p_d}{2}.
\]
Therefore, we choose the "+" sign.

For small values of \( p_3^* \) (\((p_3^*)^2 \ll M^2 + p_\perp^2\)), we expand the square root in the expression for \( p_3 \) as a Taylor series:

\[
\sqrt{(p_3^*)^2 + M^2 + p_\perp^2} \approx \sqrt{M^2 + p_\perp^2} + \frac{(p_3^*)^2}{2\sqrt{M^2 + p_\perp^2}}
\]

Substituting into the original expression:

\[
p_3 \approx \frac{E_d}{M_d} \left( p_3^* + \frac{p_d}{E_d} \left( \sqrt{M^2 + p_\perp^2} + \frac{(p_3^*)^2}{2\sqrt{M^2 + p_\perp^2}} \right) \right)
\]

%\subsection*{Спрощений результат}
\begin{align*}
p_3 &\approx \underbrace{\frac{p_d}{M_d} \sqrt{M^2 + p_\perp^2}}_{\text{leading term}} 
+ \underbrace{\frac{E_d}{M_d} p_3^*}_{\text{linear correction}} 
+ \underbrace{\frac{p_d}{2M_d} \frac{(p_3^*)^2}{\sqrt{M^2 + p_\perp^2}}}_{\text{quadratic correction}} \\
&= \frac{1}{M_d} \left( p_d \sqrt{M^2 + p_\perp^2} + E_d p_3^* + \frac{p_d (p_3^*)^2}{2\sqrt{M^2 + p_\perp^2}} \right),
\end{align*}

When $p_3^* = 0$ and $p_\perp = 0$, the longitudinal momentum equals its average value, $p_3 = p_d/2$.

However, since we are interested in the relationship between the relative momentum of the proton pair in the anti-laboratory frame and the laboratory frame (i.e., how much it deviates from $p_d/2$ in the laboratory frame), then:

\[
d p_3 \approx  \frac{1}{M_d} \left( E_d p_3^* + \frac{p_d (p_3^*)^2}{2\sqrt{M^2 + p_\perp^2}} \right)
\]

\section{Appendix B}
\label{appendix:B}
\addcontentsline{toc}{section}{Appendix B}
The complex conjugate of the unbound state:

\[
\psi_{\mathbf{k}}^*(\mathbf{r}) = D^* e^{-i \mathbf{k} \cdot \mathbf{r}} - C^* \left( \tilde{\varphi}_s^*(\mathbf{k}) \psi_s^*(\mathbf{r}) + \tilde{\varphi}_d^*(\mathbf{k}) \psi_d^*(\mathbf{r}) \right).
\]

Substituting into the normalization integral:

\[
\int \psi_{\mathbf{k}}^*(\mathbf{r}) \psi_{\text{bound}}(\mathbf{r}) \, d^3r = \int \left[ D^* e^{-i \mathbf{k} \cdot \mathbf{r}} - C^* \left( \tilde{\varphi}_s^*(\mathbf{k}) \psi_s^*(\mathbf{r}) + \tilde{\varphi}_d^*(\mathbf{k}) \psi_d^*(\mathbf{r}) \right) \right] \left( \sqrt{N_s} \psi_s(\mathbf{r}) + \sqrt{N_d} \psi_d(\mathbf{r}) \right) d^3r.
\]
Let us separately consider the two terms into which the integral can be decomposed:
\subsection*{First Term}

\[
D^* \int e^{-i \mathbf{k} \cdot \mathbf{r}} \left( \sqrt{N_s} \psi_s(\mathbf{r}) + \sqrt{N_d} \psi_d(\mathbf{r}) \right) d^3r.
\]
Since the corresponding integrations over $d^3r$ of \(\tilde{\varphi}_s(\mathbf{k})\) and \(\tilde{\varphi}_d(\mathbf{k})\) are Fourier transforms:

\[
\int e^{-i \mathbf{k} \cdot \mathbf{r}} \psi_s(\mathbf{r}) \, d^3r = (2\pi)^{3/2} \tilde{\varphi}_s(\mathbf{k}), \quad \int e^{-i \mathbf{k} \cdot \mathbf{r}} \psi_d(\mathbf{r}) \, d^3r = (2\pi)^{3/2} \tilde{\varphi}_d(\mathbf{k}).
\]
Then:
\[
D^* (2\pi)^{3/2} \left( \sqrt{N_s} \tilde{\varphi}_s(\mathbf{k}) + \sqrt{N_d} \tilde{\varphi}_d(\mathbf{k}) \right).
\]
\subsection*{Second Term}

\[
-C^* \int \left( \tilde{\varphi}_s^*(\mathbf{k}) \psi_s^*(\mathbf{r}) + \tilde{\varphi}_d^*(\mathbf{k}) \psi_d^*(\mathbf{r}) \right) \left( \sqrt{N_s} \psi_s(\mathbf{r}) + \sqrt{N_d} \psi_d(\mathbf{r}) \right) d^3r.
\]

We expand using orthogonality:

\[
= -C^* \left[ \tilde{\varphi}_s^*(\mathbf{k}) \int \psi_s^*(\mathbf{r}) \left( \sqrt{N_s} \psi_s(\mathbf{r}) + \sqrt{N_d} \psi_d(\mathbf{r}) \right) d^3r + \tilde{\varphi}_d^*(\mathbf{k}) \int \psi_d^*(\mathbf{r}) \left( \sqrt{N_s} \psi_s(\mathbf{r}) + \sqrt{N_d} \psi_d(\mathbf{r}) \right) d^3r \right].
\]
Let us compute the integrals:

\[
\int \psi_s^*(\mathbf{r}) \left( \sqrt{N_s} \psi_s(\mathbf{r}) + \sqrt{N_d} \psi_d(\mathbf{r}) \right) d^3r = \sqrt{N_s},
\]
\[
\int \psi_d^*(\mathbf{r}) \left( \sqrt{N_s} \psi_s(\mathbf{r}) + \sqrt{N_d} \psi_d(\mathbf{r}) \right) d^3r = \sqrt{N_d}.
\]

Thus:
\[
-C^* \left( \tilde{\varphi}_s^*(\mathbf{k}) \sqrt{N_s} + \tilde{\varphi}_d^*(\mathbf{k}) \sqrt{N_d} \right).
\]

Substituting into the orthogonality condition, we obtain the total expression (the complete integral):

\[
D^* (2\pi)^{3/2} \left( \sqrt{N_s} \tilde{\varphi}_s(\mathbf{k}) + \sqrt{N_d} \tilde{\varphi}_d(\mathbf{k}) \right) - C^* \left( \tilde{\varphi}_s^*(\mathbf{k}) \sqrt{N_s} + \tilde{\varphi}_d^*(\mathbf{k}) \sqrt{N_d} \right) = 0.
\]
To ensure orthogonality, if \(\tilde{\varphi}_s(\mathbf{k})\) and \(\tilde{\varphi}_d(\mathbf{k})\) are real functions, it is required that:
\[
C = (2\pi)^{3/2} D.
\]

\section*{Appendix CI. Normalization Condition of the Continuous Spectrum for Non-zero \( \mathbf{k} - \mathbf{k}' \)}

As in \cite{Kobushkin2008}, considering the relationship between the normalization constants \( C = (2\pi)^{3/2} D \) derived in Appendix B \ref{appendix:B}, let us examine the unbound wave function of the final state \( \psi_k(\mathbf{r}) = D \left( e^{i\mathbf{k}\cdot\mathbf{r}} - (2\pi)^{3/2} \psi_s(\mathbf{r}) \frac{\tilde{\phi}_s(\mathbf{k})}{N_s} \right) \). This function must satisfy the normalization condition of the continuous spectrum on the delta function:

\begin{equation}
\langle \psi_{\mathbf{k}'} | \psi_{\mathbf{k}} \rangle = (2\pi)^3 \delta^{(3)}(\mathbf{k} - \mathbf{k}')
\end{equation}
Let us explicitly expand this expression:
\begin{align}
\langle \psi_{\mathbf{k}'} | \psi_{\mathbf{k}} \rangle &= |D|^2 \int \left( e^{-i\mathbf{k}'\cdot\mathbf{r}} - (2\pi)^{3/2} \psi_s^*(\mathbf{r}) \frac{\tilde{\phi}_s^*(\mathbf{k}')}{N_s^*} \right) \nonumber \\
&\quad \times \left( e^{i\mathbf{k}\cdot\mathbf{r}} - (2\pi)^{3/2} \psi_s(\mathbf{r}) \frac{\tilde{\phi}_s(\mathbf{k})}{N_s} \right) d^3\mathbf{r}
\end{align}
We expand the brackets:
\begin{align}
&= |D|^2 \Bigg[ \int e^{i(\mathbf{k} - \mathbf{k}')\cdot\mathbf{r}} d^3\mathbf{r}  - (2\pi)^{3/2} \frac{\tilde{\phi}_s^*(\mathbf{k}')}{N_s} \int \psi_s^*(\mathbf{r}) e^{i\mathbf{k}\cdot\mathbf{r}} d^3\mathbf{r} \nonumber \\
&\quad - (2\pi)^{3/2} \frac{\tilde{\phi}_s(\mathbf{k})}{N_s^*} \int \psi_s(\mathbf{r}) e^{-i\mathbf{k}'\cdot\mathbf{r}} d^3\mathbf{r} + (2\pi)^3 \frac{ \tilde{\phi}_s^*(\mathbf{k}') \tilde{\phi}_s(\mathbf{k}) }{|N_s|^2} \int |\psi_s(\mathbf{r})|^2 d^3\mathbf{r} \Bigg]
\end{align}

The first integral gives the three-dimensional delta function:
\begin{equation}
\int e^{i(\mathbf{k} - \mathbf{k}')\cdot\mathbf{r}} d^3\mathbf{r} = (2\pi)^3 \delta^{(3)}(\mathbf{k} - \mathbf{k}')
\end{equation}

The second and third integrals are the Fourier transforms of \( \psi_s(\mathbf{r}) \):

\begin{align}
\int \psi_s^*(\mathbf{r}) e^{i\mathbf{k}\cdot\mathbf{r}} d^3\mathbf{r} &= (2\pi)^{3/2} \tilde{\phi}_s^*(\mathbf{k}) \\
\int \psi_s(\mathbf{r}) e^{-i\mathbf{k}'\cdot\mathbf{r}} d^3\mathbf{r} &= (2\pi)^{3/2} \tilde{\phi}_s(\mathbf{k}')
\end{align}

The fourth integral (normalization of \( \psi_s \)):

\begin{equation}
\int |\psi_s(\mathbf{r})|^2 d^3\mathbf{r} = N_s
\end{equation}

\subsection*{Substitution of Results}
Substituting all the integrals into the normalization condition, we obtain (and one can always substitute \( N_s = 0.98 \) into the given expression):

\begin{align}
\langle \psi_{\mathbf{k}'} | \psi_{\mathbf{k}} \rangle &= |D|^2 \Big[ (2\pi)^3 \delta^{(3)}(\mathbf{k} - \mathbf{k}') - (2\pi)^3 \frac{\tilde{\phi}_s^*(\mathbf{k}') \tilde{\phi}_s^*(\mathbf{k})}{N_s} \nonumber \\
&\quad - (2\pi)^3 \frac{\tilde{\phi}_s(\mathbf{k}) \tilde{\phi}_s(\mathbf{k}')}{N_s} + (2\pi)^3 \frac{\tilde{\phi}_s^*(\mathbf{k}') \tilde{\phi}_s(\mathbf{k}) }{|N_s|} \Big]
\end{align}

Next, for simplicity, let us consider the single-Gaussian parameterization \cite{Tartakovsky2005} (in the case of the multi-Gaussian parameterization, the conclusions remain the same). The corresponding wave function in momentum space for this parameterization is given by:

\[
\tilde{\phi}_s(\mathbf{k}) =\sqrt{N_s} \left( \frac{2}{\pi} \right)^{3/4} \alpha_i^{-3/4} e^{-k^2/(4\alpha_i)}= B e^{-k^2/(4\alpha_i)} = \tilde{\phi}_s^*(\mathbf{k}),
\]

where $0 < \alpha_i \ll 1$ is a parameter of the Gaussian width, and $B = \sqrt{N_s} \left( \frac{2}{\pi} \right)^{3/4} \alpha_i^{-3/4}$. Since the expression is real-valued, complex conjugation leaves it unchanged.

Substituting $\tilde{\phi}_s(\mathbf{k})$ and the analogous $\tilde{\phi}_s(\mathbf{k}')$ into the scalar product, we obtain:

\begin{align*}
\langle \psi_{\mathbf{k}'} | \psi_{\mathbf{k}} \rangle &= |D|^2 (2\pi)^3 \Big[
\delta^{(3)}(\mathbf{k} - \mathbf{k}') 
+ B^2 e^{-k'^2/(4\alpha_i)} e^{-k^2/(4\alpha_i)} \\
&\quad - B^2 e^{-k'^2/(4\alpha_i)} e^{-k^2/(4\alpha_i)} 
- B^2 e^{-k'^2/(4\alpha_i)} e^{-k^2/(4\alpha_i)} 
\Big] \\
&= |D|^2 (2\pi)^3 \left[
\delta^{(3)}(\mathbf{k} - \mathbf{k}') 
- B^2 e^{-(k^2 + k'^2)/(4\alpha_i)}
\right] \\
&= |D|^2 (2\pi)^3 \left[
\delta^{(3)}(\mathbf{k} - \mathbf{k}') 
-N_s \left( \frac{2 }{\pi} \right)^{3/2} \alpha_i^{-3/2} e^{- \frac{k^2 + k'^2 + 2kk' -2kk'}{4\alpha_i}}
\right].
\end{align*}

Since in the parametrization \cite{Tartakovsky2005} $\alpha_i \ll 1$, let us perform the crucial limiting procedure: Consider $\alpha_i \to 0$. Then the exponential term behaves as a delta-function, but depends on two variables. We obtain:

\[
\lim_{\alpha_i \to 0} \left[ \left( \frac{2}{\pi} \right)^{3/2} \alpha_i^{-3/2} e^{- \frac{(k-k')^2}{4\alpha_i}} \right] = \lim_{\alpha_i \to 0} \left[ 8 (2)^{3/2} \cdot \frac{1}{(4\pi \alpha_i)^{3/2}} e^{- \frac{(k-k')^2}{4\alpha_i}} \right] = 8 (2)^{3/2} \cdot \delta^{(3)}(\mathbf{k-k'}).
\]
Substituting this into the normalized scalar product condition for wave functions (at $\mathbf{k}' = \mathbf{k}$):
\[
\langle \psi_{\mathbf{k}} | \psi_{\mathbf{k}} \rangle  = |D|^2 (2\pi)^3 \left[ \delta^{(3)}(\mathbf{0}) - 8 N_s (2)^{3/2}\delta^{(3)}(\mathbf{0})  e^{- \frac{k^2}{2\alpha_i}} \right] =(2\pi)^3 \delta^{(3)}(\mathbf{0}) .
\]
where \( \delta^{(3)}(\mathbf{0}) \) is formally infinity, reflecting the normalization in the continuous spectrum.
From here, we derive the normalization factor \( |D|^2 \):
\[
|D|^2 =  \frac{ 1}{1 -8 N_s (2)^{3/2}  e^{- \frac{k^2}{2\alpha_i}} }
\]

In our approximation where $\alpha_i \to 0$, then $D \to 1$ for non-zero $k^2$ (and consequently, for non-zero $\mathbf{k} - \mathbf{k}'$).

\section*{Appendix $\mathbf{CII}$. Normalization Condition for the Continuous Spectrum at  $\mathbf{k} - \mathbf{k}' \sim 0$}

Let us examine the conditions for approximate normalization of the unbound state $\psi_{\mathbf{k}}(\mathbf{r})$ to the delta-function:

\[
\int \psi_{\mathbf{k}}^*(\mathbf{r}) \psi_{\mathbf{k}'}(\mathbf{r}) \, d^3r =(2\pi)^3 \delta(\mathbf{k} - \mathbf{k}')
\]
:
Substituting the expression for $\psi_{\mathbf{k}}(\mathbf{r})$ into this normalization condition:
\[
\psi_{\mathbf{k}}(\mathbf{r}) = D e^{i \mathbf{k} \cdot \mathbf{r}} - C \left( \tilde{\varphi}_s(\mathbf{k}) \psi_s(\mathbf{r}) + \tilde{\varphi}_d(\mathbf{k}) \psi_d(\mathbf{r}) \right)
\]

Let us expand the integral into terms:

\subsection*{Term with Plane Waves}

\[
|D|^2 \int e^{i(\mathbf{k}' - \mathbf{k}) \cdot \mathbf{r}} d^3r = |D|^2 (2\pi)^3 \delta(\mathbf{k} - \mathbf{k}')
\]

\subsection*{Cross Terms (Plane Wave $\times$ Bound States)}

\[
- D^* C \tilde{\varphi}_s^*(\mathbf{k}) \int e^{-i\mathbf{k} \cdot \mathbf{r}} \psi_s(\mathbf{r}) e^{i\mathbf{k}' \cdot \mathbf{r}} \, d^3r + \text{analogously for } \psi_d
\]
These terms vanish due to the orthogonality of plane waves and localized states.

\subsection*{Terms with Bound States}

\[
|C|^2 \left[ \tilde{\varphi}_s^*(\mathbf{k}) \tilde{\varphi}_s(\mathbf{k}') \int \psi_s^*(\mathbf{r}) \psi_s(\mathbf{r}) d^3r + \text{analogous terms}  \right] =|C|^2 \left( \tilde{\varphi}_s^*(\mathbf{k}) \tilde{\varphi}_s(\mathbf{k}') + \tilde{\varphi}_d^*(\mathbf{k}) \tilde{\varphi}_d(\mathbf{k}') \right)
\]

In deriving this expression, the normalization conditions for $\psi_s$ and $\psi_d$ were used.

From this we see that in order to satisfy the normalization condition:
\[
|D|^2 (2\pi)^3 \delta(\mathbf{k} - \mathbf{k}') + |C|^2 \left( \tilde{\varphi}_s^*(\mathbf{k}) \tilde{\varphi}_s(\mathbf{k}') + \tilde{\varphi}_d^*(\mathbf{k}) \tilde{\varphi}_d(\mathbf{k}') \right) = (2\pi)^3 \delta(\mathbf{k} - \mathbf{k}')
\]

two conditions must be fulfilled:
\begin{enumerate}
\item The coefficient in front of the delta function must be equal to 1:
\[
|D|^2 (2\pi)^3 = (2\pi)^3 \Rightarrow D = 1
\]

\item The second term, ~$|C|^2 \left( \tilde{\varphi}_s^*(\mathbf{k}) \tilde{\varphi}_s(\mathbf{k}') + \tilde{\varphi}_d^*(\mathbf{k}) \tilde{\varphi}_d(\mathbf{k}') \right)$, must vanish.
\end{enumerate}

However, this is only possible if:
\begin{itemize}
\item $C = 0$ (which contradicts the previous results, since $C = (2\pi)^{3/2} D = (2\pi)^{3/2}$; see Appendices $\mathbf{A}$, $\mathbf{B}$, and $\mathbf{CI}$),
\item or the Fourier transforms of the bound states satisfy a special condition: ~$\tilde{\varphi}_s^*(\mathbf{k}) \tilde{\varphi}_s(\mathbf{k}') + \tilde{\varphi}_d^*(\mathbf{k}) \tilde{\varphi}_d(\mathbf{k}') \ll \delta(\mathbf{k} - \mathbf{k}')$,
which holds for localized states (i.e., $\varphi_s$, $\varphi_d$ decay rapidly as $|\mathbf{k}| \to \infty$). Clearly, this condition is also satisfied for $\mathbf{k} - \mathbf{k}' = 0$.
\end{itemize}

Thus, the system of conditions is consistent under the following parameter values:
\[
D = 1, \quad C = (2\pi)^{3/2}
\]

\section{Appendix D}
\label{appendix:D}
Due to the considerable complexity of the obtained formulas, we present here, as an example, only the expression $(2\pi)^3\frac{d^3\sigma}{d^3k} $ in (\ref{eq:4.24})
for the K2 potential in the case of the s-wave
(the number of wave-function components is $N=12$, and $N_r$ is the normalization factor).

\begin{doublespace}
\begin{multline}
(2\pi)^3\frac{d^3\sigma}{d^3k} = \frac{1}{128 N_r^4 N_s^2 \pi \beta_N^2} p_d^2 (1+\rho_N^2)\sigma_N^2 
\Bigg(
N_r^2 N_s^2 (1+\rho_N^2)\sigma_N^2
\left(
\sum_{j=1}^{12}
\frac{
e^{-\frac{(2k_z+Q_z)^2 + 4(k_x^2+k_y^2+(2k_z+Q_z)^2)\beta_N^2\lambda_j}
{16\lambda_j(1+4\beta_N^2\lambda_j)}}A_j
}{
\sqrt{\lambda_j}(1+4\beta_N^2\lambda_j)
}
\right)^2
\\
+\,2N_rN_s\pi\sigma_N
\left(
\sum_{j=1}^{12}
\frac{
e^{-\frac{(2k_z+Q_z)^2+4(k_x^2+k_y^2+(2k_z+Q_z)^2)\beta_N^2\lambda_j}
{16\lambda_j(1+4\beta_N^2\lambda_j)}}A_j
}{
\sqrt{\lambda_j}(1+4\beta_N^2\lambda_j)
}
\right)
\\
\times
\Bigg[
-32N_rN_s\beta_N^2
\sum_{j=1}^{12}
\frac{
e^{-\frac{12(4k_x^2+4k_y^2+(-2k_z+Q_z)^2)\beta_N^2+\frac{(-2k_z+Q_z)^2}{\lambda_j}}
{16(1+12\beta_N^2\lambda_j)}}A_j
}{
\sqrt{\lambda_j}(1+12\beta_N^2\lambda_j)
}
\\
-\sqrt{2}
\left(
\sum_{i=1}^{12}
\frac{
e^{-\frac{k_x^2+k_y^2+k_z^2}{4\lambda_i}}A_i
}{
4N_r\sqrt{2\pi}\lambda_i^{3/2}
}
\right)
\Bigg(
(1+\rho_N^2)\sigma_N
\sum_{l=1}^{12}\sum_{m=1}^{12}
\frac{
A_l A_m e^{-\frac{Q_z^2}{16(\lambda_l+\lambda_m)}}
}{
\sqrt{\lambda_l+\lambda_m}\left(1+4\beta_N^2(\lambda_l+\lambda_m)\right)
}
\\
-32\pi\beta_N^2
\sum_{l=1}^{12}\sum_{m=1}^{12}
\frac{
A_l A_m e^{-\frac{Q_z^2}{16(\lambda_l+\lambda_m)}}
}{
\sqrt{\lambda_l+\lambda_m}\left(1+12\beta_N^2(\lambda_l+\lambda_m)\right)
}
\Bigg)
\Bigg]
\\
+\,2\pi^2
\Bigg[
64N_r^2N_s^2\beta_N^2
\sum_{j=1}^{12}
\frac{
e^{-\frac{4k_x^2+4k_y^2+(-2k_z+Q_z)^2}{16\lambda_j}}A_j
}{
\sqrt{\lambda_j}
}
\sum_{n=1}^{12}
\frac{
e^{\frac{1}{16}\left(
-\frac{4k_z^2}{\lambda_n}
+\frac{4k_zQ_z}{\lambda_n}
-\frac{Q_z^2}{\lambda_n}
-\frac{4(k_x^2+k_y^2)(-\lambda_j+16\beta_N^2\lambda_j^2-\lambda_n)}
{\lambda_j(\lambda_n+\lambda_j(1+16\beta_N^2\lambda_n))}
\right)}A_n
}{
\sqrt{\lambda_n}\left(\lambda_n+\lambda_j(1+16\beta_N^2\lambda_n)\right)
}
\\
+32\sqrt{2}N_rN_s\beta_N^2
\left(
\sum_{i=1}^{12}
\frac{
e^{-\frac{k_x^2+k_y^2+k_z^2}{4\lambda_i}}A_i
}{
4N_r\sqrt{2\pi}\lambda_i^{3/2}
}
\right)
\Bigg(
\sigma_N
\sum_{j=1}^{12}
\frac{
e^{-\frac{12(4k_x^2+4k_y^2+(-2k_z+Q_z)^2)\beta_N^2+\frac{(-2k_z+Q_z)^2}{\lambda_j}}
{16(1+12\beta_N^2\lambda_j)}}A_j
}{
\sqrt{\lambda_j}(1+12\beta_N^2\lambda_j)
}
\\
\times
\sum_{l=1}^{12}\sum_{m=1}^{12}
\frac{
A_lA_m e^{-\frac{Q_z^2}{16(\lambda_l+\lambda_m)}}
}{
\sqrt{\lambda_l+\lambda_m}(1+4\beta_N^2(\lambda_l+\lambda_m))
}
-2\pi
\sum_{l=1}^{12}\sum_{m=1}^{12}
\frac{
A_lA_m e^{-\frac{Q_z^2}{16(\lambda_l+\lambda_m)}}
}{
(\lambda_l+\lambda_m)^{3/2}
}
\\
\times
\sum_{j=1}^{12}
\frac{
e^{\frac{1}{16}\left(
-\frac{(-2k_z+Q_z)^2}{\lambda_j}
-\frac{4k_x^2(1+16\beta_N^2(\lambda_l+\lambda_m))}{
\lambda_j+\lambda_l+\lambda_m+16\beta_N^2\lambda_j(\lambda_l+\lambda_m)}
-\frac{4k_y^2(1+16\beta_N^2(\lambda_l+\lambda_m))}{
\lambda_j+\lambda_l+\lambda_m+16\beta_N^2\lambda_j(\lambda_l+\lambda_m)}
\right)}A_j(\lambda_l+\lambda_m)
}{
\sqrt{\lambda_j}
\left(
\lambda_j+\lambda_l+\lambda_m+16\beta_N^2\lambda_j(\lambda_l+\lambda_m)
\right)
}
\\
+\sum_{l=1}^{12}\sum_{m=1}^{12}\sum_{n=1}^{12}
\frac{
A_lA_mA_n
e^{\frac{1}{16}\left(
-\frac{4k_z^2}{\lambda_n}
+\frac{4k_zQ_z}{\lambda_n}
-\frac{Q_z^2}{\lambda_n}
-\frac{4(k_x^2+k_y^2)(1+16\beta_N^2(\lambda_l+\lambda_m))}
{\lambda_l+\lambda_m+\lambda_n+16\beta_N^2(\lambda_l\lambda_n+\lambda_m\lambda_n)}
\right)}
}{
\sqrt{\lambda_n}
\left(
\lambda_l+\lambda_m+\lambda_n+16\beta_N^2(\lambda_l\lambda_n+\lambda_m\lambda_n)
\right)
\sqrt{\lambda_l+\lambda_m}
}
\Bigg)
\Bigg]
\\
+\left(
\sum_{i=1}^{12}
\frac{
e^{-\frac{k_x^2+k_y^2+k_z^2}{4\lambda_i}}A_i
}{
4N_r\sqrt{2\pi}\lambda_i^{3/2}
}
\right)^2
\Bigg[
(1+\rho_N^2)\sigma_N^2
\left(
\sum_{l=1}^{12}\sum_{m=1}^{12}
\frac{
A_lA_m e^{-\frac{Q_z^2}{16(\lambda_l+\lambda_m)}}
}{
\sqrt{\lambda_l+\lambda_m}(1+4\beta_N^2(\lambda_l+\lambda_m))
}
\right)^2
\\
-64\pi\beta_N^2\sigma_N
\sum_{l=1}^{12}\sum_{m=1}^{12}
\frac{
A_lA_m e^{-\frac{Q_z^2}{16(\lambda_l+\lambda_m)}}
}{
\sqrt{\lambda_l+\lambda_m}(1+4\beta_N^2(\lambda_l+\lambda_m))
}
\sum_{l=1}^{12}\sum_{m=1}^{12}
\frac{
A_lA_m e^{-\frac{Q_z^2}{16(\lambda_l+\lambda_m)}}
}{
\sqrt{\lambda_l+\lambda_m}(1+12\beta_N^2(\lambda_l+\lambda_m))
}
\\
+128\pi^2\beta_N^2
\sum_{l=1}^{12}\sum_{m=1}^{12}\sum_{o=1}^{12}\sum_{p=1}^{12}
\frac{
A_lA_mA_oA_p e^{-\frac{Q_z^2}{16(\lambda_l+\lambda_m)}}e^{-\frac{Q_z^2}{16(\lambda_o+\lambda_p)}}
}{
(\lambda_l+\lambda_m)^{3/2}(\lambda_o+\lambda_p)^{3/2}
\left(
16\beta_N^2+\frac{1}{\lambda_l+\lambda_m}+\frac{1}{\lambda_o+\lambda_p}
\right)
}
\Bigg]
\Bigg)
\end{multline}
\end{doublespace}

\bibliographystyle{unsrt}
%\begin{thebibliography}{9}

%\begin{thebibliography}{99}

%\bibitem{landau} L. D. Landau, E. M. Lifshitz, Course of Theoretical Physics. Statistical Physics;

%\end{thebibliography}
\clearpage
\makeatletter
\@ifundefined{endNoHyper}{}{\endNoHyper}
\makeatother
\end{NoHyper}
\end{document}